\documentclass[twocolumn,pra,twocolumn,superscriptaddress]{revtex4}
\usepackage{color}
\usepackage{amsmath,amsthm,mathrsfs,dsfont}
\usepackage{graphicx}
\usepackage{tikz}
\usepackage{xr}
\usepackage{xc}
\usepackage{float}
\usepackage{bbm}
\usepackage{epsfig} 
\usepackage{verbatim}
\usepackage{array}
\usepackage{setspace}
\usepackage{bbding}
\usepackage{amssymb}% http://ctan.org/pkg/amssymb
\usepackage{pifont}% http://ctan.org/pkg/pifont
\usepackage{braket}
\usepackage{ulem}
\usepackage{newtxtext,newtxmath}

\newcolumntype{M}[1]{>{\centering\arraybackslash}m{#1}}
\newcolumntype{N}{@{}m{0pt}@{}}

\usepackage[unicode=true,
 bookmarks=true,bookmarksnumbered=false,bookmarksopen=false,
 breaklinks=false,pdfborder={0 0 1},backref=false,colorlinks=true]
 {hyperref}
\hypersetup{
 linkcolor=magenta, urlcolor=blue, citecolor=blue, pdfstartview={FitH}, hyperfootnotes=false, unicode=true}
 
\usepackage{color}
\makeatletter
\@ifundefined{textcolor}{}
{%
 \definecolor{BLACK}{gray}{0}
 \definecolor{WHITE}{gray}{1}
 \definecolor{RED}{rgb}{1,0,0}
 \definecolor{GREEN}{rgb}{0,1,0}
 \definecolor{BLUE}{rgb}{0,0,1}
 \definecolor{CYAN}{cmyk}{1,0,0,0}
 \definecolor{MAGENTA}{cmyk}{0,1,0,0}
 \definecolor{YELLOW}{cmyk}{0,0,1,0}
}

\usepackage{amsfonts}\usepackage{tabularx}\usepackage{dcolumn}\usepackage{bm}\usepackage{graphicx}\usepackage{epstopdf}

\DeclareMathOperator{\tr}{tr}

\hypersetup{urlcolor=blue}

\makeatother

\newcolumntype{C}[1]{>{\centering\arraybackslash$}p{#1}<{$}}

\usepackage{algorithm}
\usepackage{algorithmic}

\begin{document}

\widetext

\title{Scaling of entangling-gate errors in large ion crystals}

\author{Wenhao He}
\thanks{These two authors contributed equally}
\affiliation{Center on Frontiers of Computing Studies, Peking University, Beijing 100871, China}

\author{Wenhao Zhang}
\thanks{These two authors contributed equally}
\affiliation{Center on Frontiers of Computing Studies, Peking University, Beijing 100871, China}

\author{Xiao Yuan}
\email{xiaoyuan@pku.edu.cn}
\affiliation{Center on Frontiers of Computing Studies, Peking University, Beijing 100871, China}
\affiliation{School of Computer Science, Peking University, Beijing 100871, China}

\author{Yangchao Shen}
\email{shenyangchao@gmail.com}
\affiliation{Center on Frontiers of Computing Studies, Peking University, Beijing 100871, China}

\author{Xiao-Ming Zhang}
\email{xmzhang93@pku.edu.cn}
\affiliation{Center on Frontiers of Computing Studies, Peking University, Beijing 100871, China}
\affiliation{School of Computer Science, Peking University, Beijing 100871, China}

\begin{abstract} 
Trapped-ion has shown great advantages in building quantum computers. While high fidelity entangling-gate has been realized for few ions, how to maintain the high fidelity for large scale trapped-ions still remains an open problem. 
 Here, we present an analysis on arbitrary scale ion chain and focus on motional-related errors, reported as one of the leading error sources in state-of-the-art experiments. We theoretically analyze two-qubit entangling-gate infidelity in a large ion crystal. To verify our result, we develop an efficient numerical simulation algorithm that avoids exponential increases of the Hilbert space dimension. 
  For the motional heating error, We derive a much tighter bound of gate infidelity than previously estimated $O(N\Gamma\tau)$, and we give an intuitive understanding from the trajectories in the phase space of motional modes.
Our discoveries may inspire the scheme of pulse design against incoherent errors and shed light on the way toward constructing scalable quantum computers with large ion crystals.
\end{abstract}
\maketitle

\section{introduction}

Trapped-ion quantum computing, since the original proposal more than two decades ago~\cite{cirac1995quantum,solano1999deterministic,molmer1999multiparticle,sorensen1999quantum}, has shown many advantages: high entangling-gate fidelity exceeding 99.9\%~\cite{ballance2016high,gaebler2016high,clark2021high}, long coherence time beyond one hour~\cite{wang2021single}, all-to-all qubit connectivity~\cite{linke2017experimental}, and state-preparation-and-measurement fidelity over 99.99\%~\cite{Honeywell2022SPAM}. All these features make trapped-ion one of the most promising candidates for building large-scale quantum computers~\cite{kielpinski2002architecture,zhu2006trapped,zhu2006arbitrary,blatt2008entangled,monroe2013scaling,lekitsch2017blueprint,monroe2021programmable}. The ions are typically placed in a single Paul trap with interactions between different ions' spin states(two hyperfine levels~\cite{choi2014optimal}) mediated by collective motional modes and external laser fields~\cite{cirac1995quantum,solano1999deterministic} (or microwave fields with gradient~\cite{microwave_static,microwave_oscillating}). With ion's spin states defined as qubits, we can introduce entangling gate operation with specific laser pulse modulation. Molmer and Sorensen first proposed Molmer-Sorensen (MS) gate~\cite{molmer1999multiparticle,sorensen1999quantum}, which utilizes one motional mode of the ion chain. Later, a more general sheme was proposed\cite{zhu2006arbitrary} where all motional modes are considered. In this paper, we focus on this scheme, and refer to it as entangling gate.  

In recent experiments, the number of ions with fully programmable manipulation has reached double digits, both in multi short-ion-chain~\cite{Pino2021} and a single long-ion-chain ~\cite{wright2019benchmarking, pogorelov2021compact}. These technological advances fully demonstrate the scalability of ion trap systems. Obviously, for the multi short-ion-chain architecture (also known as QCCD)~\cite{kielpinski2002architecture,metodi2005quantum}, adding more ions in the operating zone can reduce the computational overhead caused by ion transport. Therefore, it becomes urgent and meaningful to address the question of how the entangling-gate error scales when we put more ions in a single trap. In this paper, we analyze and investigate the entangling-gate error with leading experimental noise theoretically and numerically,
and compare them for different ion number. Moreover, our detailed error analysis is also essential for the scheme of gate pulse design in large-scale ion-trap quantum computers in the presence of incoherent noise.

Based on the experimental results with high entangling-gate fidelity reported before~\cite{ballance2016high,gaebler2016high,clark2021high,honeywell_lightshift,wang2020high, Srinivas2021}, the leading error sources of the entangling-gate can be mainly characterized as the following types: 
\textbf{(1) spin-related}, including spontaneous emission, state decay, and spin dephasing noise; \textbf{(2) motional-modes related}, including motional heating, dephasing, and mode frequency drift; \textbf{(3) fluctuation of driving fields}, including amplitude, frequency, phase, and duration; \textbf{(4) model errors}, including pulse design imperfection, Lamb-Dicke and rotating wave approximations. 
With all motional modes mediating interactions between ion spin states, entangling gate may be influenced by motional heating in a more complicated way when ion number scales up. Hence, we focus on (2), especially the large ion number case.

 Motional heating and dephasing are typically induced by electric field noise~\cite{brownnutt2015ion,morigi2001two,home2011normal}. Ref~\cite{haddadfarshi2016high} has first provided analytical results expressions about the gate fidelity under heating errors for two-ion systems. They have also minimized the average distance to optimize resilience against heating. Ref~\cite{sutherland2022one} further provided analytical results for constant Rabi frequency and generalized the discussion to other error sources. Although its effect has been studied in details for two-ion case, the scenario for large ion number cases has rarely been studied. For multi-ion cases, one of the complications is that the pulse should be modulated in order to decouple the spin and phonon modes at the end of the gate operation. A simple estimation based on failure rate may give an infidelity scaling under motional heating $1-F=O(N\Gamma\tau)$~\cite{wu2018noise}, where $\Gamma$ is the heating rate and $\tau$ is the gate time. However, the contribution of different motional modes may be different, and the decoherence effect also depends on the control pulse shape, which makes the problem complicated. So the simple estimation cannot fully characterize the heating noise effect.  Similar to motional heating noise, parameter fluctuations also depend on the pulse shape. A detailed study of corresponding error scaling is important for scalability analysis and pulse design.

In this work, we present a detailed theoretical and numerical analysis on the motional-modes related errors of large trapped-ion systems. Based on unitary transformation, we derive an upper bound on the gate infidelity with motional-modes related errors, which depends on the trajectory of motional modes in the phase space. We then develop an efficient classical algorithm to simulate the Lindblad master equation describing the time evolution of noisy trapped-ion systems. The algorithm overcomes the exponential explosion of the Hilbert space by considering the commutation relation between different motional modes. Our numerical results show that the improved infidelity upper bound is much more accurate than the simple bound. In particular, we have considered (1) the combination of correlated and uncorrelated noise model, and (2) the fully uncorrelated noise model. In both cases, the trajectory based bound is much more accurate compared to the simple bound based on failure rate. We also show that the error depends not only on the operation time and the heating rate, but also on the Rabi frequency. We also study the errors due to motional frequency drift. In the small error region, the infidelity is proportional to the square of the parameter fluctuation noise level, which is more benign than the motional heating noise level that has linear relation to the infidelity.

The paper is organized as follows. In Section \ref{sec:Model}, we review the realization of the entangling gate for ion crystal system. In Section \ref{sec:heating}, we analyze motional-modes related error. The error bounds are compared with the numerical results obtained by an efficient simulation algorithm we develop. 
In Section \ref{sec:fluctuation}, we analyze the parameter fluctuation error both theoretically and numerically. 
In Section \ref{sec:discussion}, we provide further discussions and conclude our results.

\section{model of entangling gates} % Unnumbered section
\label{sec:Model}

\begin{figure}[t]
    \centering
       %\subfloat{%
        %  \includegraphics[height=4cm]{Figures_1a.pdf}%/Fig5_15us.png}%
        %  \label{fig:left}%
       %} 
       %\subfloat{%
        \includegraphics[width=\columnwidth]{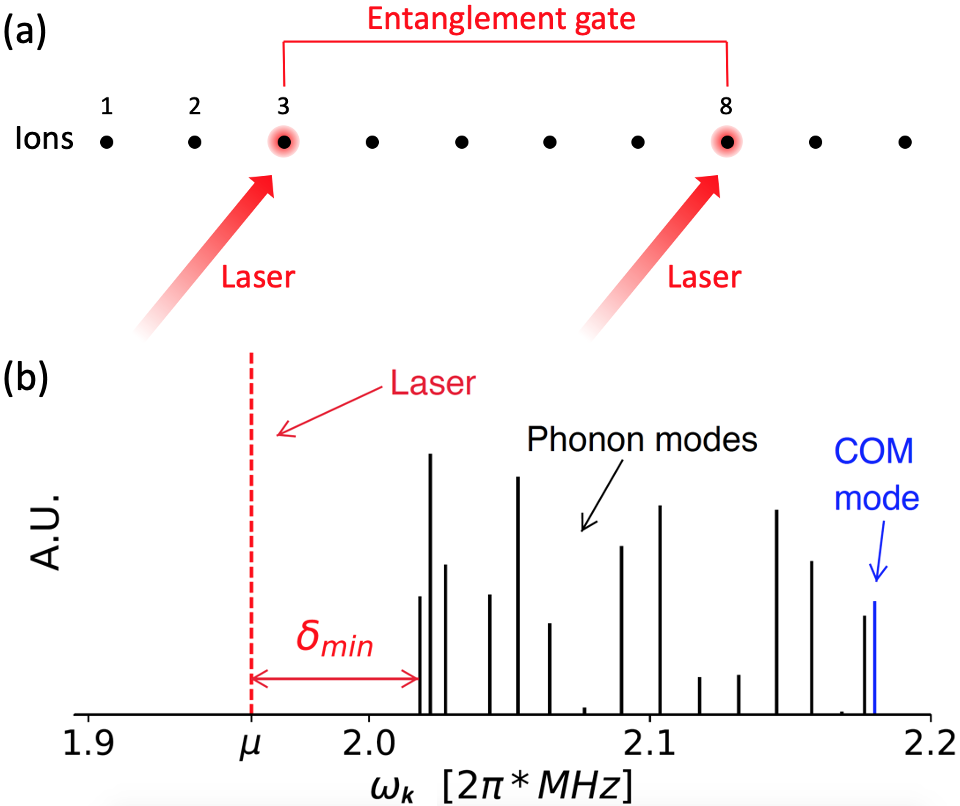}%/Fig4_15us.png}%
          \label{fig:middle}%
       %}
       \caption{Sketch of 1D ion chain. (a) All ions are aligned on a line; tuned laser beams are targeted onto two of these ions to implement a two-qubit entangling gate. (b) Effective frequency of the laser is set to be red detuned from $\min {\omega_k}$. Note the center-of-mass (COM) mode, illustrated by the blue line, has the highest frequency. The heights of the solid lines indicate the coupling strengths between the particular ion pair and all the phonon modes. $\delta_{\min}$ is defined as the detuning between the lowest phonon mode frequency and the effective laser frequency. 
       }
       \label{fig:default}
\end{figure}

We consider 1-D ion crystals consisting of $N$ ions linearly aligned along $z$ axis. The collective motion of the ions has totally $N$ transversal modes in $x$ direction, with their frequency denoted as $\omega_k$($k=1,2,...,N$). Interaction between spins and motional modes is introduced by external laser fields. As shown in Fig.~\ref{fig:default}(a), two ions, denoted as $j_a, j_b$, are coupled by shining lasers with wave vectors $k_{\text{vec}}$. Hamiltonian of the system can be described by the displacement-dependent fields applied to the spins\cite{zhu2006trapped}. 
\begin{align}
H_0(t)=\sum_{j}\Omega_j(t)\cos\left[\mu t-k_{\text{vec}}\hat x_j(t)-\phi_{j}^{(m)}\right]\sigma_j^x,\label{eq:h1}
\end{align}
where $\Omega_j(t)$ and $\mu$ are respectively the effective Rabi frequency and the frequency of the laser field at the $j$th ion, $\sigma_j^x$ is the Pauli-$X$ operator for the $j$th spin, $\phi_{j}^{(m)}$ is the phase of the laser, which is set to be 0 here and after, and $\hat x_j$ represents the quantized displacement of the $j$th spin in the $x$ direction, which can be expressed as the combination of the collective motional modes~\cite{leibfried2003quantum}
\begin{align}
\hat x_j(t)=\sum_{k}b_j^k\sqrt{\frac{\hbar}{2m_{\text{ion}}\omega_k}}(a_ke^{-i\omega_kt}+a_k^\dag e^{i\omega_kt}).\label{eq:xj}
\end{align}
Here, $m_{\text{ion}}$ is the mass of each ion, $a_k$ ($a_k^\dag$) is the annihilation (creation) operator of the $k$th mode, and $b^k_{j}\sqrt{\frac{\hbar}{2 m_{\mathrm{ion}} \omega_k}}$ characterizes the coupling strength between the $k$th motional mode and the $j$th spin, and $b^k_j$ satisfies normalization condition $\sum_{k}|b^k_{j}|^2=1$ for all $k$ and $j$. We illustrate the magnitude of $b^k_{j_a}b^k_{j_b}$ for all $k$ in Fig.~\ref{fig:default}(b). Within Lamb-Dicke regime, Eq.~\eqref{eq:h1} can be approximated as
\begin{align}
H_0 (t) \approx  \sum_{j,k}\Omega_j(t)&\sin(\mu t)\eta_{k}b_{j}^{k}(a_k^\dag e^{i\omega_kt}+a_ke^{-i\omega_kt})\sigma_j^{x},\label{eq:h2}
\end{align}
where we have neglected higher order terms about $b_j^k$ and the mode-independent single qubit rotation terms. Under the rotating-wave approximation, frequency-sum terms with factor $e^{\pm i (\mu + \omega_k) t}$ can be neglected. Because we are within Lamb-Dicke regime, $H_0(t)$ can be approximated with the Hamiltonian
\begin{align}
H(t)=\frac{i}{2}\sum_{j,k}\Omega_j(t)\eta_{k}b_{j}^{k}(a_k^\dag e^{i\delta_kt}-a_ke^{-i\delta_kt})\sigma_j^{x},\label{eq:h2}
\end{align}
where $\eta_k\equiv k_{\text{vec}}\sqrt{\hbar/2m_{\text{ion}}\omega_k}$ is the Lamb-Dick parameter of the $k$th mode, and $\delta_k=\omega_k-\mu$ is the detuning between motional modes and the laser frequency. 

The evolution under the Hamiltonian  $H(t)$ is given by~\cite{zhu2006trapped} 
\begin{align}
U(\tau)&=\exp \left[i \sum_{j} \phi_{j}^k(\tau) \sigma_{j}^{x}+i \sum_{j_1 < j_2} \Theta_{j_1,j_2}(\tau) \sigma_{j_1}^{x} \sigma_{j_2}^{x}\right], \label{eq:evo}
\end{align}
where $\phi_{j}^k(\tau)=-i\left[\alpha_{j}^{k}(\tau) a_{k}^{\dagger}-\alpha_{j}^{k*}(\tau) a_{k}\right]$ with 
\begin{align}
\alpha_{j}^{k}(\tau)=\frac{1}{2} \eta_{k} b_{j}^{k} \int_{0}^{\tau}\Omega_j(t) e^{i \delta_{k} t} \:{\rm d}t
\end{align} 
being the mode-dependent single-qubit rotation terms. During our gate design, this terms are undesired, so $\phi_j^k$ are treated as error which should be minimized. The other term $\Theta_{j_1,j_2}(\tau)$ is given by
\begin{align}
\Theta_{j_1,j_2}(\tau) = & \frac{1}{4} \sum_{k} \eta_{k}^{2} b_{j_1}^{k} b_{j_2}^{k} \int_{0}^{\tau} {\rm d} t_{1} \int_{0}^{t_{1}} {\rm d} t_{2}\: 
\big(\Omega_{j_1}(t_1)\Omega_{j_2}(t_2)\notag\\
&+\Omega_{j_2}(t_1)\Omega_{j_1}(t_2)\big) \sin \left[\delta_{k}\left(t_{1}-t_{2}\right)\right],\label{eq:theta}
\end{align}
which is the XX rotation angle. 

In this work, we focus on two qubit entangling-gates. The qubits are encoded in the spin degrees of freedom of two ions shined with the laser field (spin up $|0\rangle$ and spin down $|1\rangle$). We denote the indices of these two qubits as $j_a$ and $j_b$, and set $\Omega_{j\neq j_a,j_b}=0$. The notation of rotation angle can be simplified as $\Theta(\tau)=\Theta_{j_a,j_b}(\tau)$. In practice, we should tune the Rabi frequency such that $\Theta(\tau)$ coincides with the target rotation angle. Without loss of generality, we consider the two-qubit XX rotation gate with $\Theta(\tau)=\pi/4$ and the ideal target unitary is then
 \begin{align}
 U_{\text{id}}=\exp\left[-i\pi/4\:\sigma^{x}_{j_a}\sigma^{x}_{j_b}\right]\label{eq:uid}.
 \end{align} 
 The initial spin states are assumed to be decoupled from the motional modes, and so as the target states. Therefore, it is required that $|\alpha_j^k|=0$ for all $k$ and $j\in\{j_a,j_b\}$.
  In this work, we use  amplitude modulation to minimize $|\alpha_j^k|$, although we believe that the main results could be generalized to other modulation methods. With fixed $\mu$, a simple way to approximately get Eq.~\eqref{eq:uid} using Eq.~\eqref{eq:evo} is to modulate $\Omega_j(t)$ to minimize $|\alpha_j^k(\tau)|$ and $|\Theta-\pi/4|$. However, this method is sensitive to the fluctuation of $\mu$ and $\omega_k$. Ref.~\cite{leung2018robust} developed robust optimization protocols for frequency modulation, which has been generalized to amplitude modulations in Ref.~\cite{kang2021batch}. The main idea is that $|\alpha_j^k(\tau)|$ can be made insensitive to the fluctuation of $\mu$ to the first order with a symmetric pulse and minimizing the time average of $\alpha_j^k(t)$. More specifically, with 
\begin{subequations}\label{eq:am}
 \begin{align}
 &\Omega_j(t)=\Omega_j(\tau-t)\label{eq:amomg}\\
 &\int_0^\tau \alpha_j^k(t)\:{\rm d}t=0,\label{eq:alpha0}
 \end{align}
 \end{subequations}
 we have $|\alpha_j^k(\tau)|=0$ and 
 \begin{align}
 {\rm d}\alpha_j^k(\tau)/{\rm d}\mu=0,\quad {\rm d}\alpha_j^k(\tau)/{\rm d}\omega_k=0. %\quad d\alpha_j^k(\tau)/d\Omega_j=0.
 \end{align} 
 We also note that $\alpha_j^k(\tau)$ is insensitive to the time-independent drift of Rabi frequency as elaborated in Appendix.~\ref{sec:pf}. Therefore, when optimizing the time-dependent Rabi frequency, we use $\int_0^\tau \alpha_j^k(t){\rm d}t$ as the loss function while keeping its symmetry. In our simulation, the pulse sequences optimization problem is  transformed to a special case of quadratically constrained quadratic program~\cite{grzesiak2020efficient, blumel2021power}. %Details are provided in Appendix~\ref{ap:amp}.
After deriving a pulse sequence minimizing $\int_0^\tau \alpha_j^k(t)\:{\rm d}t$, the desired rotation angle $\Theta(\tau)=\pi/4$ can be obtained by simply multiplying $\Omega_j(t)$ by an appropriate factor.

\section{Motional-modes related error}
\label{sec:heating}

\subsection{Theoretical analysis}\label{sec:mt}
In principle, all motional modes are involved in the phonon-mediated spin-spin interaction, so the contributing error from all modes should be taken into consideration. The quantum evolution under these errors can be described by the following Lindbladian master equation~\cite{wang2020high}
\begin{align}
    \frac{\partial \rho(t)}{\partial t} = -i [H(t),\rho(t)] + \mathbb{L}(\rho(t)),
    \label{eq:mst}
\end{align}
with Hamiltonian $H(t)$ given by Eq.~(\ref{eq:h2}) and Lindbladian superoperator $\mathbb{L}$ given by
\begin{align}
\mathbb{L}(\rho) =  &\sum_k\Gamma_{k,\uparrow} \left(a_{k}^{\dagger} \rho\left(t\right) a_{k}-\frac{1}{2}\left\{a_{k} a_{k}^{\dagger}, \rho\left(t\right)\right\}\right)\notag \\
&+\sum_k\Gamma_{k,\downarrow} \left(a_{k} \rho\left(t\right) a_{k}^{\dagger}-\frac{1}{2}\left\{a_{k}^{\dagger} a_{k}, \rho\left(t\right)\right\} \right),\notag\\
&+\sum_k\Gamma_{k,\text{d}} \left(n_{k} \rho\left(t\right) n_{k}-\frac{1}{2}\left\{n_{k}^2, \rho\left(t\right)\right\} \right) \label{eq:lind},
\end{align}
where $\Gamma_{k,\text{d}}$ denotes the dephasing rate of the motional mode $k$, and $n_{k}=a_k^\dag a_k$. $\Gamma_{k,\uparrow}$ and $\Gamma_{k,\downarrow}$ correspond to the motional heating errors of the motional mode $k$. The excitation rate $\Gamma_{k,\uparrow}$ represents the average phonon number increasing per second, and the relaxation rate $\Gamma_{k,\downarrow}$ represents the average phonon number decreasing per second. If the environment is modeled with a thermal bath, a stronger coupling between the bath and systems corresponding to a larger heating rate. Moreover, $\Gamma_{k,\uparrow}$ increases with the temperature of the bath, and $\Gamma_{k,\uparrow}$ vanishes when the temperature is zero.
The heating rate values for different motional modes depend not only on the electric field noise strength on the position of each ion, but also on their spatial correlations~\cite{brownnutt2015ion}. A detailed analysis of the effect of motional heating errors will be given later in this section.

In estimating the effect of motional heating, we apply the following approximation. The first one is the small error region assumption. Let $\rho=\rho(\tau)$ be the noisy final state according to Eq.~\eqref{eq:mst}, and  $\rho^{\text{hf}}$ be the ideal heating-error-free final state generated by Eq.~\eqref{eq:mst} without the Lindbladian terms. We assume that the trace norm between them $\|\rho-\rho^{\text{hf}}\|$ is small.
Another assumption we adopt is the separable assumption, i.e. the initial spin and motional modes are disentangled with
$
\rho(0)=\rho_{\text{spin}}(0)\otimes\rho_{1}(0)\otimes\cdots\otimes\rho_{N}(0),
$
where $\rho_{\text{spin}}(0)$ is initial spin state and $\rho_{i}(0)$ is the initial state of the $i$th motional mode, which is typically in a thermal state of low average phonon number. 
We note that while analytical results below are guaranteed with the assumptions above, they may hold in more general scenarios. 

The error is characterized by the infidelity between the final states generated with and without Lindbladian terms. Because we are only interested in the spin subspace, the motional modes are traced out after the evolution. More specifically, we characterize the motional-modes related error with
\begin{equation}
    1-F(\tr_{\text{ph}}(\rho),\tr_{\text{ph}}(\rho^{\text{hf}})),
    \label{eq:hea_err_def}
\end{equation}
Here, $F(\sigma_1, \sigma_2)=(\operatorname{tr} \sqrt{\sqrt{\sigma_1} \sigma_2 \sqrt{\sigma_1}})^{2}$ is the fidelity between two states. Note that $\rho^{\text{hf}}$ is expected to be close to the target state $U_{\text{id}}\rho_{\text{spin}}(0)U_{\text{id}}^\dag$ if the amplitude modulation is well-behaved.

One way to estimate Eq.~(\ref{eq:hea_err_def}) is based on the property of the fidelity function under partial trace $F(\rho,\rho^{\text{hf}})<F(\tr_{\text{ph}}\rho,\tr_{\text{ph}}\rho^{\text{hf}})$~\cite{nielsen2002quantum,nielsen1996entanglement}. By considering only the first two levels of each mode~\footnote{One may perform energy cut-off with at a higher level. But this will not change the scaling, of the bound. Moreover, the bound will become evel looser.}, and evaluating the failure probability, one may obtain a simple error bound (details in Appendix \ref{sec:math}) as
\begin{equation}
1-F(\tr_{\text{ph}}\rho,\tr_{\text{ph}}\rho^{\text{hf}}) \leqslant 1 - F(\rho,\rho^{\text{hf}}) \leqslant  \sum_k (\Gamma_{k,\uparrow}+\Gamma_{k,\downarrow}+\frac{\Gamma_{k,\text{d}}}{4}) \tau,
\label{eq:old_bnd}
\end{equation}
where $\tau$ is the gate duration. This bound is consistent with the estimation provided in the  literature, such as~\cite{wu2018noise}. However, it treats all motional modes equally, and ignores an important fact that only few motional modes have significant contribution to the phonon-meditated spin-spin interaction process. So there are rooms for improving the estimation about the motional-modes related errors. Indeed, it has been reported that Eq.~(\ref{eq:old_bnd}) is larger than numerical values by several magnitudes~\cite{wu2018noise}.

Here, we present a more elaborate analysis of the master equation Eq.~(\ref{eq:mst}). With a transformation which effectively eliminates the unitary evolution part of Eq.~(\ref{eq:mst}), we obtain an improved error bound (see Appendix \ref{sec:math} for details)
\begin{align}
    &1-F(\tr_{\text{ph}}\rho,\tr_{\text{ph}}\rho^{\text{hf}}) \nonumber\\
    \leqslant& \sum_{j_1,j_2\in\{j_a,j_b\}} \left\lvert \sum_{k}(\Gamma_{k,\uparrow}+\Gamma_{k,\downarrow}+\Gamma_{k,d})\int_{0}^{\tau} {\rm d} t\: \alpha_{j_1}^{k*}(t) \alpha_{j_2}^{k}(t)\right\rvert \nonumber \\ 
    +&O(\Lambda^2+A^4), \label{eq:tight_bnd}
\end{align}
where $O(\Lambda^2)$ is the higher order term related to $\|\rho-\rho^{\text{hf}}\|$, which is negligible in the small error region (see Appendix~\ref{sec:math}). Moreover, $A=\sum_{j_1,j_2\in\{j_a,j_b\}}\sum_{k}\int_{0}^{\tau} {\rm d} t\: |\alpha_{j_1}^{k*}(t) \alpha_{j_2}^{k}(t)|$, and $O(A^4)$ is the higher order terms related to the trajectory in the phase space, which is neglected in our analysis. Eq.~\eqref{eq:tight_bnd} can be interpreted as follows. In the phase space, $|\int_{0}^{\tau} d t \alpha_{j_1}^{k*}(t) \alpha_{j_2}^{k}(t)|$ is proportional to the integration of the distance to the ground state. The motional-modes related error becomes larger when the state of motional modes are far away from the ground state. Numerical simulation presented in the next section and Appendix~\ref{ap:sup} shows that Eq.~\eqref{eq:tight_bnd} is much tighter than Eq.\eqref{eq:old_bnd}. Moreover, Eq.~\eqref{eq:tight_bnd} may still be valid for various models of  $\Gamma_{k,\mu}$.

We can further simplify Eq. \eqref{eq:tight_bnd} as a relatively looser bound (see Appendix.~\ref{sec:math} for details) as
\begin{align}\label{eq:loose_bnd}
1-F(\tr_{\text{ph}}\rho,\tr_{\text{ph}}\rho^{\text{hf}})&\leqslant \max_{k,j\in\{j_a,j_b\}}(\Gamma_{k,\uparrow}+\Gamma_{k,\downarrow}+\Gamma_{k,\text{d}}) \notag\\
 &\times\eta_k^2 \int_{0}^{\tau} {\rm d} t \left\lvert \int_{0}^{t} {\rm d} t_{1}\: \Omega_{j}\left(t_{1}\right) e^{-i \omega_{k} t_{1}}\right\rvert ^2\notag\\
 &+O(\Lambda^2+A^4).
\end{align}
From Eq. \eqref{eq:loose_bnd}, we can derive an infidelity scaling as 
\begin{equation}\label{eq:scl}
    1-F(\tr_{\text{ph}}\rho,\tr_{\text{ph}}\rho^{\text{hf}}) =O(\Omega^2_{\text{max}}\eta \Gamma_{\max}\tau^3)
\end{equation}
where $\eta = \max_k \eta_k$, $\Gamma_{\max} = \max_{k,\beta\in\{\uparrow,\downarrow\}} \{\Gamma_{k,\beta}\}$, $\Omega_{\text{max}} = \max_{j,t}\{ \Omega_j(t)\}$, and $\tau$ is the gate duration. 

We also note that there is still room for improving Eq.~\eqref{eq:scl}, especially for the dependency on $\tau$. Our numerical results below (Fig.~\ref{fig:infid_tau}) show that the error increase much slower than $O(\tau^3)$. A tighter bound can be achieved if we impose more restrictions on the pulse, and take the initial state into consideration. 

\begin{algorithm} [H]
\caption{\texttt{sequential mode simulation}\label{alg:alg1} }  
\begin{algorithmic}[1]

\STATE \textbf{Input} : 
\STATE \quad initial disentangled state $\rho_{\text{spin}}(0)\otimes\rho_1(0)\otimes...\otimes\rho_N(0)$,

\STATE \quad Hamiltonian $\sum_k H_k$ and Lindbladian $\sum_k \mathbb{L}_k$, 

\STATE \quad gate duration $\tau$

\STATE \textbf{For} $k = 1\cdots N$ \textbf{do}: 

\STATE \quad $\rho' \leftarrow \rho'_{\text{spin}} \otimes \rho_k(0)$;

\STATE \quad $d\rho'/dt=-i[H_k,\rho']+\mathbb{L}_k(\rho')$ from time $t=0$ to $t=\tau$;

\STATE \quad $\rho'_{\text{spin}} \leftarrow \tr_{\text{ph}}(\rho') $;

\STATE \textbf{Output} $\rho'_{\text{spin}} $;

\end{algorithmic} 
\end{algorithm} 

\subsection{Numerical simulation}
\label{sec:num_sim}

The direct simulation of Eq.~\eqref{eq:mst} is intractable. For an ion-chain system composed of $N$ ions, there are totally $N$ motional modes in the $x$ direction. During two-qubit gates implementation, all the motional modes are  entangled with the spin states of both ions. So the full Hilbert space has dimension $4\times N_{\text{c}}^{N}$, where $N_{\text{c}}$ is the cut-off dimension of the Fock space for each mode. So the simulation complexity scales exponentially with $N$. Traditional numerical approaches for open systems, such as Monte Carlo wave-function method~\cite{molmer1993monte} would also suffer from this problem. One possible method to solve this problem is to keep only few modes into consideration~\cite{wu2018noise,wang2020high,sorensen1999quantum}, but it will result in a significant compromise of the simulation accuracy, especially when the ion number is large.

By considering the commutation relation between spin and phonon degree of freedoms, we developed an efficient simulation algorithm with linear runtime (Algorithm.~\ref{alg:alg1}). The main idea is that within the Lamb-Dicke regime, the dissipation effect on each motional mode can be treated sequentially. More details are provided in Appendix.~\ref{app:seq}.

We are focusing on two-qubit entangling gates, so the spin subspace is $4$-dimensional. Therefore, during each iteration, Algorithm~\ref{alg:alg1} always works on a Hilbert space with $4\times N_{c}$ dimensions. Therefore, it reduces the runtime from $O(N_c^{3N})$ to $O(N\times N_c^3)$, where the cubic index comes from matrix multiplication. The simulation time grows linearly with $N$. On the other hand, with the brute-force method of solving Eq.~\eqref{eq:mst} directly, the simulation is already intractable for $N=3$ and $N_c=10$. A detailed comparison is provided  in Appendix.~\ref{ap:time}. 

\begin{figure}
	\centering
	\includegraphics[width=1\columnwidth]{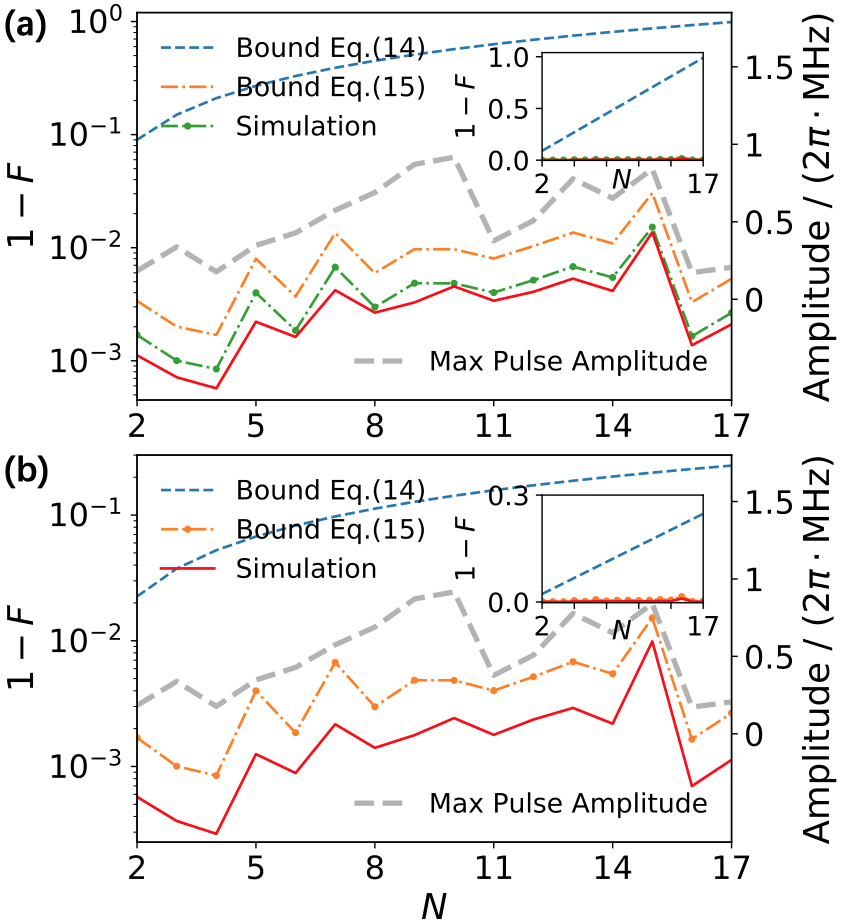}
	\caption{Upper bound and simulation results. (a) Infidelity under motional heating error with $\Gamma_{\text{COM},\uparrow}=\Gamma_{\text{COM},\downarrow}=50\times N\, \text{phonon/s}$, $\Gamma_{k\neq\text{COM},\uparrow}=\Gamma_{k\neq\text{COM},\downarrow}=50\, \text{phonon/s}$ and $\Gamma_{k,\text{d}}=0$. (b) Infidelity under dephasing error with $\Gamma_{\text{COM},\text{d}}=50\times N\, \text{phonon/s}$, $\Gamma_{k\neq\text{COM},\text{d}}=50\, \text{phonon/s}$ and $\Gamma_{k,\uparrow}=\Gamma_{k,\downarrow}=0$. Dashed line (blue) in both subfigures: simple bound given in Eq.~\eqref{eq:old_bnd}; dashed doted line (orange) in both subfigures: improved bound given in Eq.~\eqref{eq:tight_bnd}; dashed doted line with solid dot (green): estimation given in Eq.~\eqref{eq:stm} for motional-modes related error; solid line (red): simulation results. We have set $\tau=300\mu\mathrm{s}$. Gray dashed line shows the maximal pulse amplitudes under different ion numbers.
	}
	\label{fig:infid_w_bounds}
\end{figure}

With Algorithm.~\ref{alg:alg1}, we are ready to examine the theoretical analysis in Sec.~\ref{sec:mt}. In the first step, we should determine the 
 heating rate values $\Gamma_{k,\uparrow}$, $\Gamma_{k,\downarrow}$, $\Gamma_{k,\text{d}}$ in the simulation, and we try to make our noise model close to the real experimental cases. The main contribution of motional heating is the electric field fluctuation, which is in general spatially correlated. Therefore, in determining $\Gamma_{k,\mu}$, one should take not only the noise strength, but also the correlation into consideration. When noise is fully correlated, only the COM mode [blue lines in Fig.~\ref{fig:default}(b)] has non-zero heating rates, which increases linearly with $N$. On the other hand, for completely uncorrelated noise, heating rates are approximately uniform for each mode \cite{brownnutt2015ion}.  The completely correlated model may hold if there are only few ions, and the case will become more complicated when the length of ion-chain becomes comparable to the ion electrode distance. In the main text, we consider the combination of fully correlated and uncorrelated noise, and discuss the motional heating error and dephasing error separately.  More specifically, when considering motional heating error, we set $\Gamma_{\text{COM},\uparrow}=\Gamma_{\text{COM},\downarrow}=\Gamma N$, $\Gamma_{k\neq\text{COM},\uparrow}=\Gamma_{k\neq\text{COM},\downarrow}=\Gamma$ and $\Gamma_{k,\text{d}}=0$; when considering dephasing error, we set $\Gamma_{\text{COM},\text{d}}=\Gamma N$, $\Gamma_{k\neq\text{COM},\text{d}}=\Gamma$ and $\Gamma_{k,\uparrow}=\Gamma_{k,\downarrow}=0$. 
 
\begin{figure}
    \centering 
    \includegraphics[height = 10.2cm]{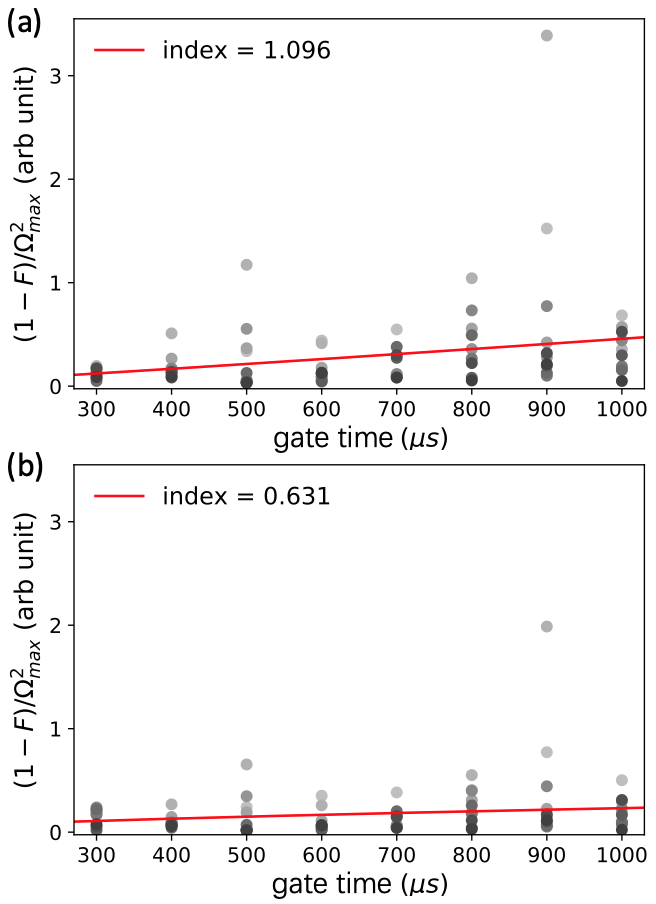}
       \caption{
       Rescaled infidelity $(1-F)/\Omega_{\max}^2$ versus gate time $\tau$. 
       Each dot represents a specific gate time $\tau$ and laser frequency. 
       We set $N=17$, $\delta_{\min}=2\pi\times\{0.01{\rm MHz},0.02{\rm MHz},\cdots,0.10{\rm MHz}\}$, 
       and the opacity of the dots increases with $\delta_{\min}$. 
       Heating rate is set as  $\Gamma_{k\neq\mathrm{COM}} = 50$ phonons$/{\rm s}$, $\Gamma_{\mathrm{COM}} = 50\times N$ phonons$/{\rm s}$.
       (a) Results under motional heating errors.
       (b) Results under dephasing errors. Red lines represent the fitting with $y=k\tau^p$, where $y$ is the average value of rescaled infidelity.  We obtain $p=1.096$ and $p=0.631$ for (a) and (b) respectively. We realized the best $\delta_{\min}$ yields an error level far better than averaged on all $\delta_{\min}$. For heating error the ratio ranges from 2.2 to 9.1 and for dephasing error 2.4 to 10.9. This result suggests possibilities of optimizing the pulse to thermaliztion.}
       \label{fig:infid_tau}
\end{figure} 

\begin{figure}[t]
    \centering
          \includegraphics[height = 7.2cm]{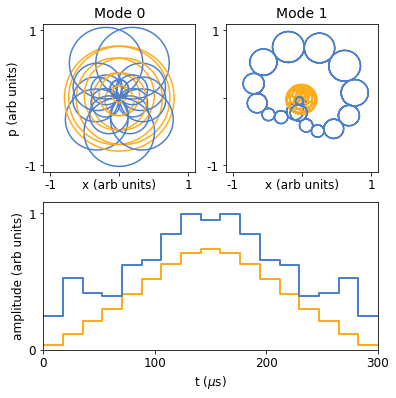}
                 
       \caption{Trajectory in phase space and  pulse sequence for $N=2$.  Blue: $\delta_{\min}=2\pi\times0.03$MHz with infidelity $0.33\times10^{-3}$. Orange: $\delta_{\min}=2\pi\times0.06$MHz with infidelity $1.12\times10^{-3}$. (a) and (b) demonstrate trajectories of motional modes 0 and 1, respectively. (c) demonstrate the absolute value of Rabi frequency  versus time. We have only considered the motional heating error, i.e. $\Gamma_{k,\text{d}}=0$, $\Gamma_{0,\uparrow}=\Gamma_{0,\downarrow}=50$ photons$/$s, $\Gamma_{1,\uparrow}=\Gamma_{1,\downarrow}=100$ photons$/$s. }
       \label{fig:pulse}%
\end{figure} 

We first examine the validity of error bounds under motional heating errors. Fig.~\ref{fig:infid_w_bounds}(a) shows the comparison between error bounds in Eq.~\eqref{eq:old_bnd} and Eq.~\eqref{eq:tight_bnd}, and numerical simulation results. We have used the abbreviation $F\equiv F(\tr_{\text{ph}}\rho,\tr_{\text{ph}}\rho^{\text{hf}})$.
Here and after, we use $|00\rangle$ as the initial state of the spin subspace, and assume all motional modes are initialized at the ground state. It can be noticed that the fluctuation of Eq.~\eqref{eq:tight_bnd} exhibits a similar pattern to the one for simulation results, and the deviation does not change significantly with $N$. On the other hand, the simple bound of Eq.~\eqref{eq:old_bnd}, increases linearly with $N$, which is inconsistent with the numerical results. In particular, Eq.~\eqref{eq:old_bnd} is about $10^2\sim10^3$ times larger than the simulation results. In contrast, even for the worst case, Eq.~\eqref{eq:tight_bnd} is only about four times larger than the simulation result.

Besides, a more accurate estimation (instead of the upper bound) can be given by taking the population of motional modes into consideration. During evolution, most motional modes are close to the ground state. The contribution from $\Gamma_{k,\downarrow}$ is significantly lower than $\Gamma_{k,\uparrow}$. We may therefore approximate the heating error by the infidelity estimator 
\begin{align}
\sum_{j_1,j_2\in\{j_a,j_b\}} \left\lvert \sum_{k}\Gamma_{k,\uparrow}\int_{0}^{\tau} {\rm d}t\: \alpha_{j_1}^{k*}(t) \alpha_{j_2}^{k}(t)\right\rvert,\label{eq:stm}
\end{align}
 which is closer to the numerical simulation results. 
In Fig.~\ref{fig:infid_w_bounds}(b), we provide numerical results for dephasing error. Similarly, compared to Eq.~\eqref{eq:old_bnd},  the bound in Eq.~\eqref{eq:tight_bnd} is closer to the numerical results for at least an order of magnitude closer. Moreover, the fluctuation in Eq.~\eqref{eq:tight_bnd} also is similar to the one for the simulation results. We note that there is still a gap between Eq.~\eqref{eq:tight_bnd} and simulation results. One of the possible reason is that Eq.~\eqref{eq:tight_bnd} is the worst-case estimation over all possible initial states. A tighter estimation can be obtained by taking the initial-state-dependency into consideration.

In both Fig.~\ref{fig:infid_w_bounds}(a) and (b), we have also shown the maximal Rabi frequency under different ion numbers $N$ in our pulse design. There is no significant raise of the pulse amplitudes in the parameter range considered in this work . 

We then discuss the relation between infidelity and gate time $\tau$. Eq.~\eqref{eq:scl} indicates that the infidelity increase at most  cubically with $\tau$. But in practice, the fidelity may increase much slower than $O(\tau^3)$. For ion number $N=17$ with different $\delta_{\text{min}}$, we calculate the infidelity for $\tau$ ranging from $\tau=300\mu$s to $1000\mu$s. Because the optimized pulse shape have different $\Omega_{\max}$ for different $\tau$ and $\delta_{\text{min}}$, we focus on the rescaled infidelity $(1-F)/\Omega_{\max}^2$ in Fig.~\ref{fig:infid_tau}. For both motional heating error (Fig.~\ref{fig:infid_tau}(a)) and dephasing error (Fig.~\ref{fig:infid_tau}(b)), we calculate the rescaled infidelity for $\delta_{\min}$ ranging from $2\pi\times0.01{\rm MHz}$ to $2\pi\times0.1{\rm MHz}$. To verify that the infidelity scales slower than $O(\tau^3)$, we show the fitting with curve $y=k\tau^p$ (red lines), where $y$ is the averaged value of rescaled infidelity $(1-F)/\Omega_{\max}^2$ over all $\delta_{\text{min}}$. The indexes $p$ obtained by fitting is $p=1.096$ and $p=0.631$ for motional heating error and dephasing error respectively. Our numerical results indicates that the infidelity increases much slower than the upper bound $p\leqslant 3$ provided by Eq.~\eqref{eq:scl}. We also note that the difference of infidelity for different $\delta_{\text{min}}$ is significant. In particular, for $\tau=900\mu$s, the maximum of infidelity is $6.5$ times (or $34$ times) larger than the minimum. Therefore, in practical implementation, the optimization over laser frequency is necessary for minimizing the motional-modes related errors.

To have a better understanding of the relation between infidelity and pulse shape, in Fig.~\ref{fig:pulse}, we illustrate the trajectory in phase space and Rabi frequencies. In Fig.~\ref{fig:infid_w_bounds}(a), and (b), we set $N=2$, and illustrate the trajectory in phase space for $k=0$ and $k=1$ respectively. Blue and orange lines correspond to the largest infidelity ($1.12\times10^{-3}$) and smallest infidelity ($0.33\times10^{-3}$) among all choices of $\delta_{\text{min}}$, respectively. We have set $\tau=300\mu$s in order to be consistent with Fig.~\ref{fig:infid_w_bounds}, although it can be further reduced in practice. As can be seen, the trajectory for small infidelity is much closer to the origin point in the phase space, which corresponds to a smaller $|\alpha_j^k|$. This is consistent with the relation between $|\alpha_j^k(t)|$ and infidelity indicated by Eq.~\eqref{eq:tight_bnd}. In Fig.~\ref{fig:infid_w_bounds}(c), we demonstrate the absolute value of Rabi frequency at different time steps. As can be seen, the amplitude for the pulse with smaller infidelity is much smaller, which is consistent with Eq.~\eqref{eq:scl}. Our result is consistant with former observation that encompassing the same area, there is less time-averaged spin-motion entanglement for origin-centered trajectories than non-origin centered trajectories\cite{haddadfarshi2016high,shapira2018robust,webb2018resilient,sutherland2020laser}.

\begin{figure}[t]
    \centering
       \includegraphics[height=8cm]{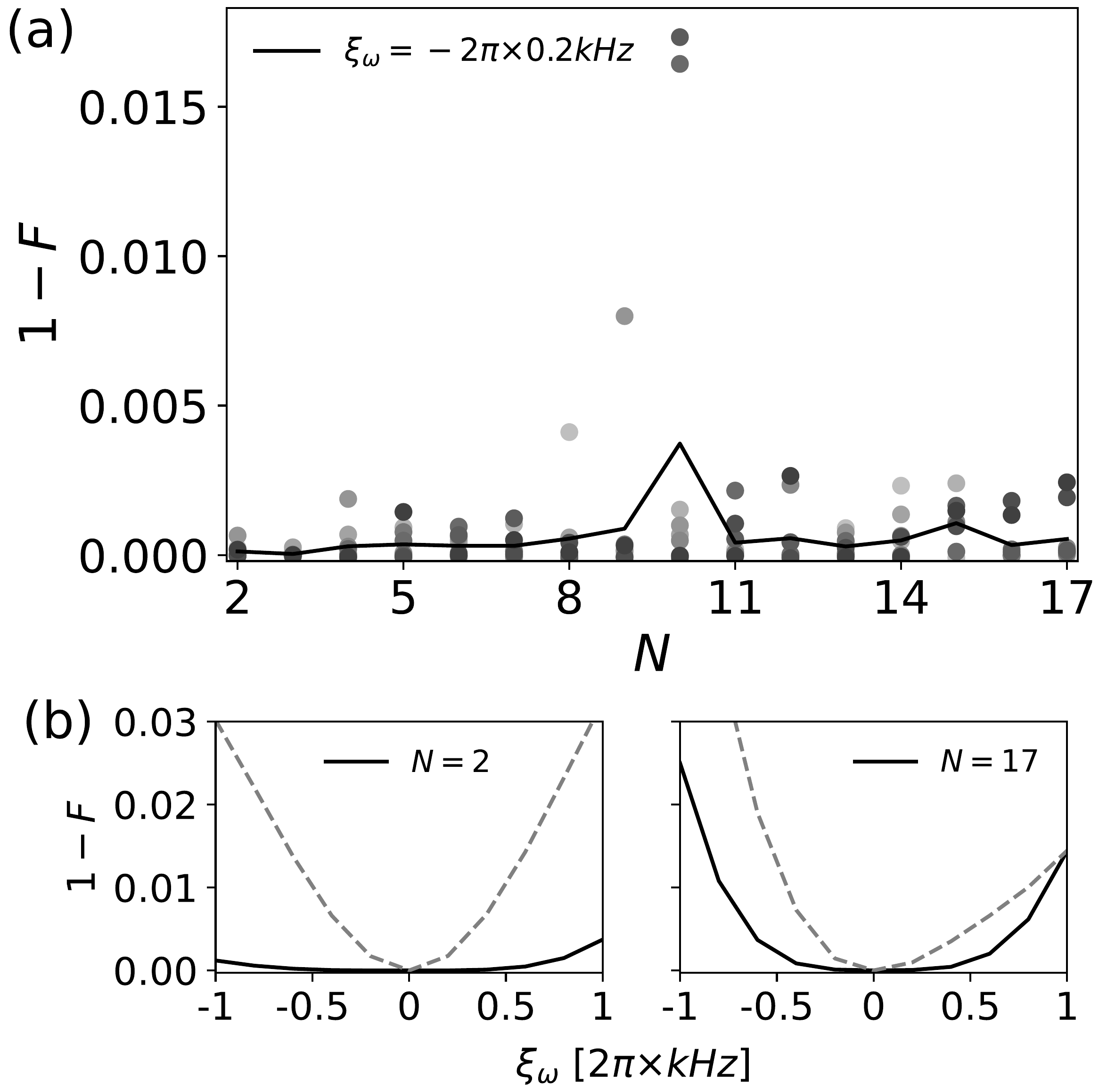}
       \caption{Infidelity under motional frequency drift. (a) infidelity versus ion number $N$ with $\xi_\omega = -2\pi\times0.2{\rm kHz}$. Gate time is fixed to be $\tau = 300{\rm \mu s}$. Each dot represents a specific laser frequency $\mu$ corresponding to the system with $N$ ions, with $\delta_{\min}=2\pi\times\{0.01{\rm MHz},0.02{\rm MHz},\cdots,0.10{\rm MHz}\}$, and the opacity of the dots increases with $\delta_{\min}$. The solid line represents the mean value over all $\delta_{\text{min}}$. (b) Infidelity versus $\xi_{\omega}$ for $N=2$  (left) and $N=17$ (right) with (solid lines) and without (dashed lines) robustness condition. } 
       \label{fig:Rabi_and_mu}
\end{figure} 

\section{Mode frequency drift}
\label{sec:fluctuation} 
 
Mode frequency drift is another important motional-mode-related error sources~\cite{leung2018robust,kang2021batch}. We consider a quasi-static fluctuation noise model, i.e. the noise strengths are fixed during gate time $\tau$. Moreover, the fluctuation of all motional modes are also assumed to be identical. More specifically, during the gate synthesis, we still apply Eq.~\eqref{eq:evo}, while the parameters are replaced by
\begin{align}\label{eq:err1}
\delta_k&\rightarrow \delta_k+\xi_\omega.
\end{align}
The first term of Eq.~\eqref{eq:evo} about $\phi_j(\tau)$ is insensitive to $\xi_\omega$ at the first order \cite{leung2018robust,kang2021batch}, so the infidelity mainly comes from the change of $\Theta(\tau)$. More specifically, the infidelity between the ideal final state and the noisy final states under fluctuations can be estimated as $1-F\leqslant (\Theta(\tau)-\tilde\Theta(\tau))^2$, where $\tilde\Theta(\tau)$ is the rotation angle $\Theta(\tau)$ with $\delta_k$ replaced in the form of Eq.~\eqref{eq:err1}. After some calculation, we obtain
\begin{align}
1-F\leqslant\Delta_\omega+O(\xi_{\omega}^4),\label{eq:inf_f}
\end{align}
where

\begin{align}
\Delta_\omega&=\xi_{\omega}^2\times\notag\\
&\left|\sum_{k} \eta_{k}^{2} b_{j_a}^{k} b_{j_b}^{k} \int_{0}^{\tau} {\rm d} t_{1} \int_{0}^{t_{1}} {\rm d} t_{2} 
\chi(t_1,t_2) \cos \left(\delta_{k}\Delta t_{1,2}\right)\Delta t_{1,2}\right|^2, \label{eq:Delta_mu}
\end{align}
and $\chi(t_1,t_2)=\Omega_{j_a}(t_1)\Omega_{j_b}(t_2)
+\Omega_{j_b}(t_1)\Omega_{j_a}(t_2)$, and $\Delta t_{1,2}=t_1-t_2$. 
The error due to the fluctuation of $\omega_k$ can be bounded by $\Delta_\omega\leqslant O(\xi_{\omega}^2 \eta^{4}\Omega_{\max}^4\tau^6)$. The infidelity is proportional to the square of the fluctuation. So compared to motional-modes related error, the infidelity is much insensitive to the mode frequency drift. We also note that in practice, $\xi_{\omega}$ may scales much slower than $O(\Omega_{\max}^4\tau^6)$ with respect to Rabi frequency and time, especially when $\Omega_{j}(t)$ varies much slower than $1/\delta_k$. 

To analysis the relation between infidelity and ion number, in Fig.~\ref{fig:Rabi_and_mu}(a), we plot the infidelity with ion number from $N=2$ to $N=17$. As can be seen, the infidelity has no obvious increase with $N$. In particular, mean value of infidelity over all $\delta_{\text{min}}$ and all ion numbers $N\in[2,17]$ is $6.4\times10^{-4}$, while the mean value for $N=17$ is $5.4\times10^{-4}$. Moreover, similar to the scenario for motional-modes related errors (Fig.~\ref{fig:infid_tau}), the infidelity varies significantly for different detuning $\delta_{\min}$. Take $N=10$ as an example, the maximum of infidelity over all $\delta_{\text{min}}$ is $0.0173$, while the minimum of infidelity is $<10^{-4}$. Therefore, in practical implementation, the detuning should be optimized for minimizing the mode frequency drift error. In Fig.~\ref{fig:Rabi_and_mu}(b) we compare the infidelity for pulse sequences optimized with and without robustness condition. The results show that the robust pulse sequences used in this work is much more insensitive to $\xi_{\omega}$. This result is consistent with existing literatures~\cite{leung2018robust,kang2021batch}.

\section{discussion and conclusion}
We have discussed the error scaling of two-qubit gates in large ion crystals. For motional-modes related errors, our works are summarized as follows: (1) we provide a pulse-specific lower bond of infidelity, Eq.~\eqref{eq:tight_bnd}, which is easy to calculate and close to the numerical results; (2) we develop an efficient simulation algorithm overcoming the exponential explosion of Hilbert space and enabling numerical estimation of the gate infidelity of large-scale trapped-ion system with a \textit{linear} runtime;
(3) base on Eq.~\eqref{eq:tight_bnd}, we derive an error scaling $O(\Omega^2_{\text{max}}\eta \Gamma_{\max}\tau^3)$ (Eq.~\eqref{eq:scl}).  

Several remarks for the pulse design with minimal gate error are as follows:

\textbf{I:} Compared to the simple estimation in Eq.~\eqref{eq:old_bnd}, Eq.~\eqref{eq:tight_bnd} has no explicit dependency on $N$. Instead, the infidelity depends on other parameters, including the operation time and the Rabi frequency. On the other hand, it is possible that the infidelity raise after one increase $N$ further. This is because there is a lower bound of $\tau$ or Rabi frequency to ensure that the solution of pulse optimization exist, and that lower bound should increase with $N$. 
%We also note that our infidelity bound does not include initial spin state, because related terms are eliminated in inequality reduction Eq.~\eqref{ap:eq:1}.

\textbf{II:} The infidelity depends on both the evolution time and Rabi frequency. So reducing the evolution time does not always lead to a smaller infidelity, because the Rabi frequency will also be increased. One may optimize the parameters as follows. For different evolution time, one optimizes the control pulse and estimate the corresponding infidelity. Then, one chooses the evolution time with minimal infidelity whose  corresponding Rabi frequency is within the reach of experiment.

%\textbf{II:} \red{The relationship between motional-modes related errors, Rabi frequency and operation time are interrelated and there is a power-time trade-off. According to our simulation, with a fixed Rabi frequency bound, the infidelity increases with gate time. Therefore, the power-time trade-off can be dealt with the following procedure. We first determine the threshold Rabi frequency. Then, from small gate time to large, we perform gate optimization to minimize $|\alpha_j^k(\tau)|$, and record the maximal Rabi frequency. We increase the gate time until maximal Rabi frequency is lower than the threshold value. In this way, we obtain the minimal gate time under a fixed threshold Rabi frequency. Alternatively, we can fix the gate time threshold, and vary the Rabi frequency with a similar procedure. }

\textbf{III:} Our results provide insights about gate optimization. According to Fig.~\ref{fig:infid_tau} and Fig.~\ref{fig:Rabi_and_mu}, the infidelity depends highly on the detuning $\delta_{\text{min}}$. So we should choose the optimal laser frequency for implementation. Moreover, infidelity difference for different laser pulse optimization settings as shown in Fig.~\ref{fig:infid_w_bounds} and App.~\ref{ap:sup} also indicate that the pulse shapes are yet to be optimized for motional-modes related errors robustness.  Eq.~\eqref{eq:tight_bnd} provides a cost function for motional-modes related errors minimization, and reducing infidelity by pulse engineering is our ongoing work.

We have also studied the  mode frequency drift error under the quasi-static model. The infidelity is proportional to the square of parameter drifts. So compared to the motional heating that depends on noise strength linearly, the quantum gate is much less vulnerable to parameter fluctuation.

There are several questions remain open. The first one is the generalization to multi-qubit gates. For multi-qubit gates, the number of constrains for pulse optimization increases linearly with the ions involved in the quantum gates. Secondly, our analysis has neglected higher order terms, which includes couplings between different motional modes, counter-roating terms, etc. Higher-order terms may be required to achieve a more accurate estimation of the infidelity, especially for ultra-fast quantum gate.

 We thank Yukai Wu and Zixuan Huo for helpful discussions. This work is supported by the National Natural Science Foundation of China Grant No. 12175003. The numerical simulation is supported by High-performance Computing Platform of Peking University.

\label{sec:discussion}

\bibliographystyle{apsrev4-1}
\bibliographystyle{abbrv}
\bibliographystyle{unsrt}
%\bibliography{sample}

\onecolumngrid
\newpage
\begin{appendix}

\section{Mathematical details of motional-modes related errors Analysis}
\label{sec:math}
\subsection{Simple error bound in Eq.~\eqref{eq:old_bnd} under two-level approximation}\label{app:simple}
A simple way of estimating the infidelity is based on the failure rate, i.e. the probability of having phonon modes being relaxed or excited. Under the noise model described by Eq.~\eqref{eq:mst}, the failure rate increases linearly with the phonon number. So we perform phonon number cut-off to avoid the infinite failure rate. Because the phonon mode are close to the ground state, in below, we only keep the first excitation of each mode. In other words, each phonon mode is treated as a two-level system.

The failure probability is linear to the operation time and the sum of heating rate for all motional modes as will be shown in below. We define $\Delta\tau=\tau/n_{\text{step}}$, $t_j = j\Delta \tau$, and $\Delta\rho_j=\rho (t_{j})-\rho^{\text{hf}} (t_{j})$, where $\rho$ and $\rho^{\text{hf}}$ is defined by
\begin{align}
    &\partial_t \rho = -i [H,\rho] + \sum_k\mathbb{L}_k(\rho) \\
    &\partial_t \rho^{\text{hf}} = -i [H,\rho^{\text{hf}}] \\
    &\rho (t=0) = \rho^{\text{hf}}(t=0)
\end{align}
respectively. At moment $t_{j+1}$, the distance between $\rho$ and $\rho^{\text{hf}}$ is
\begin{align}
	&\|\rho (t_{j+1})-\rho^{\text{hf}} (t_{j+1})\| \notag\\
	=&\left\| e^{-iH\Delta t}\left(\rho^{\text{hf}} (t_j)+\Delta\rho_{j}\right)e^{iH\Delta\tau} + \Delta\tau \sum_k\mathbb{L}_k(\rho(t_j))-\rho^{\text{hf}} (t_{j+1})\right\| +O(\Delta\tau^2)\notag \\
	 =&\left\|e^{-iH\Delta t}\Delta\rho_{j}e^{iH\Delta\tau} + \Delta\tau \sum_k\mathbb{L}_k(\rho(t_j))\right\| +O(\Delta\tau^2)\notag\\
	 \leqslant&\left\|e^{-iH\Delta t}\Delta\rho_{j}e^{iH\Delta\tau} \right\|+ \left\|\Delta\tau \sum_k\mathbb{L}_k(\rho(t_j))\right\| +O(\Delta\tau^2)\notag\\
     =& \|\rho (t_{j})-\rho^{\text{hf}} (t_{j})\| + \Delta \tau  \left\|\ \sum_k\mathbb{L}_k(\rho(t_j))\right\| +O(\Delta\tau^2)\notag
  \label{ap:eq:NGt}
\end{align}
where $\|\cdot\|$ represents the trace norm. Using this argument iteratively, we have 
\begin{align}
    \left\|\rho (\tau)-\rho^{\text{hf}} (\tau)\right\| \leqslant \tau \sum_k\max_j \left\|\ \mathbb{L}_k(\rho(t_j))\right\|
\end{align}
When phonon number is limited for every mode, $ \max_j \left\|\ \mathbb{L}_k(\rho(t_j))\right\|$ can be also bound by a constant. For example, assuming that motional mode is only limited to ground and first excitation state, $\left\|\rho (\tau)-\rho^{\text{hf}} (\tau)\right\|\leqslant \tau\sum_k\left(2\Gamma_{k,\uparrow}+2\Gamma_{k,\downarrow}+\Gamma_{k,\text{d}}/2\right)$.

So we have 
\begin{align}
1-F(\tr_{\text{ph}}\rho,\tr_{\text{ph}}\rho^{\text{hf}})\leqslant  1-F(\rho (\tau),\rho^{\text{hf}})\leqslant\frac{1}{2}\left\|\rho (\tau)-\rho^{\text{hf}} (\tau)\right\|\leqslant \tau\sum_k\left(\Gamma_{k,\uparrow}+\Gamma_{k,\downarrow}+\Gamma_{k,\text{d}}/4\right)
\end{align}

\subsection{Improved bound in Eq.~\eqref{eq:tight_bnd}}
A better error bound could be given with more careful analysis of master equation Eq.~\eqref{eq:mst}. Different from Appendix.~\ref{app:simple}, in analysis below, we does not apply the two-level approximation. We first apply a unitary transformation with unitary operator defined in Eq.~\eqref{eq:evo} with $\tau$ replaced by $t$, $\tilde{\rho}(t) = U^\dagger(t)\rho U(t)$, $\tilde{a}_k(t) = U^\dagger(t) a_k U(t)$. Master equation Eq.~\eqref{eq:mst} becomes
\begin{align}
    \frac{\partial \tilde{\rho}(t)}{\partial t} =  \tilde{\mathbb{L}}_t(\tilde{\rho}(t))
\end{align}
where
\begin{align}
\tilde{\mathbb{L}}_t(\tilde \rho) = \sum_{k}\mathbb{L}_{t,k,\uparrow}(\tilde{a}_k^\dag,\tilde \rho)+\mathbb{L}_{t,k,\downarrow}(\tilde{a}_k,\tilde \rho)+\mathbb{L}_{t,k,\text{d}}(\tilde{n}_k,\tilde \rho),
\end{align}
$\tilde{n}_k=\tilde{a}_k^\dag\tilde{a}_k$, and
\begin{align}
\mathbb{L}_{t,k,\mu}(\hat O, \rho) =  &\Gamma_{k,\mu} \left(\hat O \rho\hat O^\dag-\frac{1}{2}\left\{\hat O^\dag \hat O, \rho\left(t\right)\right\}\right).
\label{eq:lind}
\end{align}

Let $\Lambda=\int_0^\tau dt \, \tilde{\mathbb{L}}_t(\tilde{\rho}(0))$, the evolution can be linearly approximated as
\begin{align}
    \tilde{\rho}(\tau) = \tilde{\rho}(0)+ \Lambda+O(\Lambda^2).
\end{align}
We focus on a region that the error introduced by the heating noise is small, i.e. $\|\rho-\rho^{\text{hf}}\|\ll1$. This is equivalent to $\|\Lambda\|\ll1$ which enable us to neglect the higher order term of $O(\Lambda^2)$. We note that $\|\rho-\rho^{\text{hf}}\|\ll1$ can be ensured when $\sum_{k}(\Gamma_{k,\uparrow}+\Gamma_{k,\downarrow}+\Gamma_{k,\text{d}})\tau\ll1$. But our numerical analysis indicates that $\|\rho-\rho^{\text{hf}}\|\ll1$ is satisfied for wider range of parameters, even when $\sum_{k}(\Gamma_{k,\uparrow}+\Gamma_{k,\downarrow}+\Gamma_{k,\text{d}})\tau$ is larger than $1$.

Transforming back to the original picture, we have
\begin{align}
    \rho (\tau) &\approx \rho^{\text{hf}} + \int_0^\tau {\rm d}t \, \bar{\mathbb{L}}_t(\rho^{\text{hf}}),
    \end{align}
where    
\begin{align}
    \bar{a}_k(t) &=  U_0(\tau) \tilde{a}_k(t) U_0^\dagger(\tau)\label{eq:appeq2}\\
    &= a_k + \sum_{j\in \{j_a,j_b\}}(\alpha_j^k(t) - \alpha_j^k(\tau))\sigma_j^x \\
    &:= a_k + S_k(t)\\
    \bar{\mathbb{L}}_t(\rho^{\text{hf}})&=\sum_k\mathbb{L}_{t,k,\uparrow}(\bar a_k^\dag,\rho^{\text{hf}})+\mathbb{L}_{t,k,\downarrow}(\bar a_k,\rho^{\text{hf}})+\mathbb{L}_{t,k,\text{d}}(\bar n_k,\rho^{\text{hf}}),
\end{align}
and $\bar n_k=\bar a_k^\dagger\bar a_k$.
Here $\rho^{\text{hf}}$ is the heating-error-free quantum state generated without Lindbladian terms, and $S_k(t)$ is a time dependent operator only onperating on ions' spin states. In deriving Eq.~\eqref{eq:appeq2}, properties of creation and annihilation operators are exploited\cite{scully1999quantum}. The infidelity between two ideal and noisy quantum states in the spin subspace can be estimated according to the trace distance.

Distance between states generated with and without Lindbladian terms can be estimated as 
\begin{align}
    1-F(\tr_{\text{ph}}\rho,\tr_{\text{ph}}\rho^{\text{hf}}) &\leqslant \frac{1}{2} \left\lVert \text{tr}_{\text{ph}}(\rho(\tau) - \rho^{\text{hf}})\right\rVert \notag\\
    &\leqslant  \frac{1}{2} \left\lVert \text{tr}_{\text{ph}}\int_0^\tau {\rm d}t \, \left(\bar{\mathbb{L}}_{t,\uparrow}(\rho^{\text{hf}})+\bar{\mathbb{L}}_{t,\downarrow}(\rho^{\text{hf}})+\bar{\mathbb{L}}_{t,\text{d}}(\rho^{\text{hf}})\right)\right\rVert+O(\Lambda^2) \notag\\
    &\leqslant\frac{1}{2} \left\lVert \text{tr}_{\text{ph}}\int_0^\tau {\rm d}t \, \bar{\mathbb{L}}_{t,\uparrow}(\rho^{\text{hf}})\right\rVert+\frac{1}{2} \left\lVert \text{tr}_{\text{ph}}\int_0^\tau {\rm d}t \, \bar{\mathbb{L}}_{t,\downarrow}(\rho^{\text{hf}})\right\rVert+\frac{1}{2} \left\lVert \text{tr}_{\text{ph}}\int_0^\tau {\rm d}t \, \bar{\mathbb{L}}_{t,\text{d}}(\rho^{\text{hf}})\right\rVert+O(\Lambda^2),\label{ap:eq:infi}
\end{align}
where $\text{tr}_{\text{ph}}$ is the partial trace of all motional modes, and we have defined $\bar{\mathbb{L}}_{t,\uparrow}(\rho^{\text{hf}})=\sum_k\mathbb{L}_{t,k,\uparrow}(\bar a_k^\dag,\rho^{\text{hf}})$, $\bar{\mathbb{L}}_{t,\downarrow}(\rho^{\text{hf}})=\sum_k\mathbb{L}_{t,k,\downarrow}(\bar a_k^\dag,\rho^{\text{hf}})$, $\bar{\mathbb{L}}_{t,\text{d}}(\rho^{\text{hf}})=\sum_k\mathbb{L}_{t,k,\text{d}}(\bar a_k^\dag,\rho^{\text{hf}})$ as the contribution from excitation, relaxation and dephasing terms respectively. In below, we discuss the contribution from each term separately.
\subsubsection{motional heating error}
 The motional heating error contains excitation and relaxation terms. We consider the excitation term first,
\begin{align}
&\frac{1}{2}\left\lVert \text{tr}_{\text{ph}}\int_0^\tau {\rm d}t \, \bar{\mathbb{L}}_{t,\uparrow}(\rho^{\text{hf}})\right\rVert\\
    =&\frac{1}{2}\text{tr}_{\text{ph}}\left(\int_0^\tau {\rm d}t \,  \sum_k \Gamma_{k,\uparrow}\left(\bar{a}_{k}^{\dagger} \rho^{\text{hf}} \bar{a}_{k}-\frac{1}{2}\left\{\bar{a}_{k} \bar{a}_{k}^{\dagger}, \rho^{\text{hf}}\right\}\right)\right) \\
    =&\frac{1}{2}\int_0^\tau {\rm d}t\,\sum_k\Gamma_{k,\uparrow} \left( \text{tr}_{\text{ph}}\left({a}_{k}^{\dagger} \rho^{\text{hf}} {a}_{k}-\frac{1}{2}\left\{{a}_{k} {a}_{k}^{\dagger}, \rho^{\text{hf}}\right\}\right)\right. \notag\\
    &+ \text{tr}_{\text{ph}}\left({a}_{k}^{\dagger} \rho^{\text{hf}} {S}_{k}+{S}_{k}^{\dagger} \rho^{\text{hf}} {a}_{k}-\frac{1}{2}\left\{{a}_{k} {S}_{k}^{\dagger}+{S}_{k} {a}_{k}^{\dagger}, \rho^{\text{hf}}\right\}\right)\notag\\
    &\left. + \text{tr}_{\text{ph}}\left({S}_{k}^{\dagger} \rho^{\text{hf}} {S}_{k}-\frac{1}{2}\left\{{S}_{k} {S}_{k}^{\dagger}, \rho^{\text{hf}}\right\}\right)  \right).\label{eq:exc1}
\end{align}
With the property of trace operator, the first term in Eq.~\eqref{eq:exc1} equals zero and the second term can be further simplified as
\begin{align}
    &\frac{1}{2}\text{tr}_{\text{ph}}\left({a}_{k}^{\dagger} \rho^{\text{hf}} {S}_{k}+{S}_{k}^{\dagger} \rho^{\text{hf}} {a}_{k}-\frac{1}{2}\left\{{a}_{k} {S}_{k}^{\dagger}+{S}_{k} {a}_{k}^{\dagger}, \rho^{\text{hf}}\right\}\right) \\
    =&\frac{1}{2}\text{tr}_{\text{ph}}\left({a}_{k}^{\dagger} \rho^{\text{hf}}\right) {S}_{k}+\frac{1}{2}{S}_{k}^{\dagger} \tr_{\text{ph}}\left(\rho^{\text{hf}} {a}_{k}\right)-\frac{1}{4}\tr_{\text{ph}}\left(\left\{{a}_{k} , \rho^{\text{hf}}\right\}\right){S}_{k}^{\dagger}-\frac{1}{4}{S}_{k} \tr_{\text{ph}}\left(\left\{ {a}_{k}^{\dagger}, \rho^{\text{hf}}\right\}\right)\\
    =&\frac{1}{2} \left[\text{tr}_{\text{ph}}({a}_{k}^{\dagger} \rho^{\text{hf}}), {S}_{k}\right] + H.c.\\
    =&\frac{1}{2} \left[\text{tr}_{\text{ph}}\left({a}_{k}^{\dagger} U_0(\tau)\rho(0)U_0^\dagger(\tau)\right), {S}_{k}\right] + H.c.
\end{align}
Recall that we have assumed that for initial state of spin and motional modes are separable, the the motional modes are at thermal states. If the pulse is well optimized, we have $\alpha(\tau)=0$, so
\begin{align}
    \text{tr}_{\text{ph}}\left({a}_{k}^{\dagger} U(\tau)\rho(0)U^\dagger(\tau)\right) =0.%\propto tr(a^\dagger\sum_{l}c_l \ket{l}\bra{l}) = 0.
\end{align}
The remaining term is the third term. We can bound the norm of the third term as % infidelity equals the third term, norm of which can be bounded

\begin{align}
    & \frac{1}{2}\left\lVert \int_{0}^{\tau} {\rm d} t \sum_{k} \Gamma_{k,\uparrow}\operatorname{tr}_{\text{ph}}\left(S_{k}^{\dagger} \rho^{\mathrm{hf}} S_{k}-\frac{1}{2}\left\{S_{k} S_{k}^{\dagger}, \rho^{\mathrm{hf}}\right\}\right) \right\rVert\\
    =& \frac{1}{2}\left\lVert\int_{0}^{\tau} {\rm d} t \sum_{k}\Gamma_{k,\uparrow} S_{k}^{\dagger} \rho^{\mathrm{in}} S_{k}-\frac{1}{2}\left\{S_{k} S_{k}^{\dagger}, \rho^{\mathrm{in}}\right\} \right\rVert \\
    \leqslant& \frac{1}{2}\sum_{j_1,j_2\in\{j_a,j_b\}} \left\lvert \sum_{k}\Gamma_{k,\uparrow} \int_{0}^{\tau} {\rm d} t\: \alpha_{j_1}^{k*}(t) \alpha_{j_2}^{k}(t)\right\rvert \left( \lVert \sigma_{j_1}\rho^{\text{in}}\sigma_{j_2}\rVert + \frac{1}{2}\lVert\left\{\sigma_{j_2}\sigma_{j_1},\rho^{\text{in}}\right\}  \rVert \right) \label{ap:eq:1}\\
    \leqslant &\sum_{j_1,j_2\in\{j_a,j_b\}} \left\lvert \sum_{k} \Gamma_{k,\uparrow} \int_{0}^{\tau} {\rm d} t \alpha_{j_1}^{k*}(t) \alpha_{j_2}^{k}(t)\right\rvert. \label{ap:eq:2} 
    \end{align}
    which is similar to Eq.~\eqref{eq:tight_bnd} in the main text, and we have defined $\rho_{\text{in}} = \text{tr}_{\text{ph}}(\rho^{\text{hf}})$. Eq.~\eqref{ap:eq:2} can be further bounded by a simpler expression 
    \begin{subequations}\label{eq:loose_sub}
    \begin{align}
    \leqslant & \frac{1}{4}\sum_{j_1,j_2\in\{j_a,j_b\}} \sum_{k}\Gamma_{k,\uparrow} \left\lvert \eta_k^2 b_{j_1}^k b_{j_2}^k \int_{0}^{\tau} {\rm d} t \int_{0}^{t} {\rm d} t_1\:  \Omega_{j_1}(t_1)e^{-i\omega_k t_1}\int_{0}^{t} {\rm d} t_2 \: \Omega_{j_2}(t_2)e^{i\omega_k t_2}\right\rvert \\
    \leqslant & \frac{1}{4}\Gamma_{\uparrow} \sum_{j_1,j_2\in\{j_a,j_b\}} \max_k\left\lvert\eta_k^2\int_{0}^{\tau} {\rm d} t \int_{0}^{t} {\rm d} t_{1}\: \Omega_{j_{1}}\left(t_{1}\right) e^{-i \omega_{k} t_{1}} \int_{0}^{t} {\rm d} t_{2}\: \Omega_{j_{2}}\left(t_{2}\right) e^{i \omega_{k} t_{2}}\right\rvert\sqrt{\sum_k b_{j_1}^{k2} \sum_k b_{j_2}^{k2}} \label{ap:eq:3}\\
    \leqslant & \Gamma_{\uparrow} \max_{k,j_1,j_2} \left\lvert\eta_k^2\int_{0}^{\tau} {\rm d} t \int_{0}^{t} {\rm d} t_{1}\: \Omega_{j_{1}}\left(t_{1}\right) e^{-i \omega_{k} t_{1}} \int_{0}^{t} {\rm d} t_{2}\: \Omega_{j_{2}}\left(t_{2}\right) e^{i \omega_{k} t_{2}}\right\rvert\label{ap:eq:4}\\
    \leqslant & \Gamma_{\uparrow} \max_{k,j_1,j_2} \eta_k^2\sqrt{\int_{0}^{\tau} {\rm d} t \left\lvert \int_{0}^{t} {\rm d} t_{1}\: \Omega_{j_{1}}\left(t_{1}\right) e^{-i \omega_{k} t_{1}}\right\rvert ^2  \int_{0}^{\tau} {\rm d} t \left\lvert \int_{0}^{t} {\rm d} t_{2}\: \Omega_{j_{2}}\left(t_{2}\right) e^{-i \omega_{k} t_{2}}\right\rvert ^2}\label{ap:eq:5}\\
    = & \Gamma_{\uparrow} \max_{k,j\in\{j_a,j_b\}} \eta_k^2 \int_{0}^{\tau} {\rm d} t \left\lvert \int_{0}^{t} {\rm d} t_{1}\: \Omega_{j}\left(t_{1}\right) e^{-i \omega_{k} t_{1}}\right\rvert ^2 \label{ap:eq:6}
\end{align}
\end{subequations}
where $\Gamma_{\uparrow}=\max\{\Gamma_{k,\uparrow}\}$. We have used $\alpha_j^k(\tau) = 0$ in Eq.~\eqref{ap:eq:1}, used the property of trace norm in Eq.~\eqref{ap:eq:2}, used Cauchy inequality in Eq.~\eqref{ap:eq:3},~\eqref{ap:eq:5}, and used unity summation of $b_j^k$ in Eq.~\eqref{ap:eq:4}.

Similar analysis can be applied for relaxation terms. Substituting these results into inequalities above and considering that the infidelity is bounded by trace distance\cite{nielsen2002quantum,fuchs1999cryptographic}, with which we obtain
\begin{align}
\frac{1}{2}\left\lVert \text{tr}_{\text{ph}}\int_0^\tau {\rm d}t \, \bar{\mathbb{L}}_{t,\downarrow}(\rho^{\text{hf}})\right\rVert\leqslant \sum_{j_1,j_2\in\{j_a,j_b\}}\left\lvert \sum_{k} \Gamma_{k,\downarrow} \int_{0}^{\tau} {\rm d} t\: \alpha_{j_1}^{k*}(t) \alpha_{j_2}^{k}(t)\right\rvert \label{eq:relax_tight}
\end{align}
or a looser bound with simpler form Eq.~\eqref{ap:eq:6}, 
\begin{align}
    \frac{1}{2}\left\lVert \text{tr}_{\text{ph}}\int_0^\tau {\rm d}t \, \bar{\mathbb{L}}_{t,\downarrow}(\rho^{\text{hf}})\right\rVert\leqslant  \Gamma_{\downarrow} \max_{k,j\in\{j_a,j_b\}} \eta_k^2 \int_{0}^{\tau} {\rm d} t \left\lvert \int_{0}^{t} {\rm d} t_{1}\: \Omega_{j}\left(t_{1}\right) e^{-i \omega_{k} t_{1}}\right\rvert ^2 \label{ap:eq:bound},
\end{align}
where $\Gamma_{\downarrow}=\max\{\Gamma_{k,\downarrow}\}$. They are correspond to Eq.~\eqref{eq:tight_bnd} and Eq.~\eqref{eq:loose_bnd} respectively.

\subsubsection{Dephasing error}
We then consider the contribution form dephasing term.Dephasing satisfies 
\begin{align}
\frac{1}{2} \left\lVert \text{tr}_{\text{ph}}\int_0^\tau {\rm d}t \, \bar{\mathbb{L}}_{t,\text{d}}(\rho^{\text{hf}})\right\rVert=&\frac{1}{2} \left\lVert \sum_{k}\int_0^\tau {\rm d}t \,\text{tr}_{\text{ph}} \bar{\mathbb{L}}_{t,k,\text{d}}(\rho^{\text{hf}})\right\rVert.
\end{align}
 Recall that $\bar{\mathbb{L}}_{t,\text{d}}(\rho^{\text{hf}})=\sum_k\bar{\mathbb{L}}_{t,k,\text{d}}(\bar{n}_k,\rho^{\text{hf}})$, where $\bar n_k=n_k+a_kS_k^*+a^\dag_kS_k+|S_k|^2$, and $\bar{\mathbb{L}}_{t,k,\text{d}}(\bar{n}_k,\rho^{\text{hf}})=\Gamma_{k,\text{d}}\left(\bar n_k\rho^{\text{hf}}\bar n_k-\frac{1}{2}\left\{\bar n_k\bar n_k,\rho^{\text{hf}}\right\}\right)$.
Firstly, we have
\begin{align}
\text{tr}_{\text{ph}}\left(\bar n_k\rho\bar n_k\right)&=\text{tr}_{\text{ph}}\left(\left(n_k+a_kS_k^*+a^\dag_kS_k+|S_k|^2\right)\rho^{\text{hf}}\left(n_k+a_kS_k^*+a^\dag_kS_k+|S_k|^2\right)\right)\notag\\
&=\text{tr}_{\text{ph}}\left(\left(n_k+a_kS_k^*+a^\dag_kS_k+|S_k|^2\right)\rho_\text{ph}\otimes\rho_{\text{spin}}\left(n_k+a_kS_k^*+a^\dag_kS_k+|S_k|^2\right)\right)\notag\\
&=|S_k|^4\rho_{\text{spin}}+ S_k\rho_{\text{spin}}S_k^*.\label{eq:deph1}
%&=|f_k(t)|^2|1\rangle\langle1|\otimes S_x\rho_{\text{spin}}S_x+\text{non-diagonal and higher order terms }
\end{align}
Moreover,
\begin{align}
\text{tr}_{\text{ph}}\left(\frac{1}{2}\left\{\bar n_k\bar n_k,\rho^{\text{hf}}\right\}\right)&=\frac{1}{2}\text{tr}_{\text{ph}}\left(\left\{\left(n_k+a_kS_k^*+a^\dag_kS_k+|S_k|^2\right)^2,\rho^{\text{hf}}\right\}\right)\notag\\
&=\left(|S_k|^2+|S_k|^4\right)\rho_{\text{spin}}.\label{eq:deph2}
%&=\bar n_k\left(n_k+i(a_kf_k(t)-a_k^\dagger f_{k}^*(t))S_x+|f(t)|^2S_x^2\right)|0\rangle\langle0|\otimes\rho_{\text{spin}}\\
%&=\bar n_k\left(0+i(0f_k(t)-|1\rangle\langle0| f_{k}^*(t))S_x+|f(t)|^2S_x^2|0\rangle\langle0|\right)\otimes\rho_{\text{spin}}\\
%&=\bar n_k\left(-i|1\rangle\langle0| f_{k}^*(t)S_x+|f(t)|^2S_x^2|0\rangle\langle0|\right)\otimes\rho_{\text{spin}}\\
%&=\left(-i(if_k(t)|0\rangle\langle0|) f_{k}^*(t)S_x^2+|f(t)|^4S_x^4|0\rangle\langle0|\right)\otimes\rho_{\text{spin}}+\text{non-diagonal terms }\\
%&=|f_k(t)|^2|0\rangle\langle0|\otimes S_x^2\rho_{\text{spin}}+\text{non-diagonal and higher-order terms }\\
\end{align}
Note that we have assumed that phonon modes are all initialized to the ground states. Combining Eq.~\eqref{eq:deph1} with Eq.~\eqref{eq:deph2}, we have 
\begin{align}
&\frac{1}{2}\left\lVert \sum_k\int_0^\tau {\rm d}t \,\text{tr}_{\text{ph}} \bar{\mathbb{L}}_{t,k,\text{d}}(\rho^{\text{hf}})\right\rVert\notag\\
=&\frac{1}{2}\left\|\sum_k\Gamma_{k,\text{d}}\left(|S_k|^2\rho_{\text{spin}}-S_k\rho_{\text{spin}}S_k^*\right)\right\|+O\left(A^4\right)\notag\\
\leqslant&\sum_{j_1,j_2\in\{j_a,j_b\}} \left\lvert  \sum_k\Gamma_{k,\text{d}}\int_{0}^{\tau} {\rm d} t \alpha_{j_1}^{k*}(t) \alpha_{j_2}^{k}(t)\right\rvert+O\left(A^4\right),\label{eq:deph_tight}
\end{align}
where $A=\sum_{j_1,j_2\in\{j_a,j_b\}}\sum_{k}\int_{0}^{\tau} {\rm d} t\: |\alpha_{j_1}^{k*}(t) \alpha_{j_2}^{k}(t)|$ is the higher-order related to the trajectories in the phase space.
With a similar derivation to Eq.~\eqref{eq:loose_sub}, the contribution from dephasing error can be further simplified as
\begin{align}
\frac{1}{2} \left\lVert \text{tr}_{\text{ph}}\int_0^\tau {\rm d}t \, \bar{\mathbb{L}}_{t,\text{d}}(\rho^{\text{hf}})\right\rVert\leqslant&  \Gamma_{\text{d}} \max_{k,j\in\{j_a,j_b\}} \eta_k^2 \int_{0}^{\tau} {\rm d} t \left\lvert \int_{0}^{t} {\rm d} t_{1}\: \Omega_{j}\left(t_{1}\right) e^{-i \omega_{k} t_{1}}\right\rvert ^2 +O\left(A^4\right),\label{eq:deph_loose}
\end{align}
where $\Gamma_{\text{d}}=\max\{\Gamma_{k,\text{d}}\}$.

Combining Eq.~\eqref{ap:eq:infi}, Eq.~\eqref{ap:eq:2}, Eq.~\eqref{eq:relax_tight}, Eq.~\eqref{eq:deph_tight} we have 
\begin{align}
    &1-F(\tr_{\text{ph}}\rho,\tr_{\text{ph}}\rho^{\text{hf}}) \nonumber\\
    \leqslant& \sum_{j_1,j_2\in\{j_a,j_b\}} \left\lvert \sum_{k}(\Gamma_{k,\uparrow}+\Gamma_{k,\downarrow}+\Gamma_{k,d})\int_{0}^{\tau} {\rm d} t\: \alpha_{j_1}^{k*}(t) \alpha_{j_2}^{k}(t)\right\rvert \nonumber \\ &+O(\Lambda^2+A^4),
\end{align}
which is equivalent to Eq.~\eqref{eq:tight_bnd} in the main text.

Moreover, combining Eq.~\eqref{ap:eq:infi}, Eq.~\eqref{ap:eq:6},Eq.~\eqref{ap:eq:bound} and Eq.~\eqref{eq:deph_loose}, we obtain 
\begin{align}
1-F(\tr_{\text{ph}}\rho,\tr_{\text{ph}}\rho^{\text{hf}})&\leqslant \max_{k,j\in\{j_a,j_b\}}(\Gamma_{k,\uparrow}+\Gamma_{k,\downarrow}+\Gamma_{k,\text{d}}) \eta_k^2 \int_{0}^{\tau} {\rm d} t \left\lvert \int_{0}^{t} {\rm d} t_{1}\: \Omega_{j}\left(t_{1}\right) e^{-i \omega_{k} t_{1}}\right\rvert ^2+O(\Lambda^2+A^4).
\end{align}
which is equivalent to Eq.~\eqref{eq:loose_bnd} in the main text.

\section{Mathmatical details for mode frequency drift}\label{sec:pf}

We denote $\tilde U(\tau)$ as the unitary in Eq.~\eqref{eq:evo} with its parameters replaced in the form of Eq.~\eqref{eq:err1}. Other noise values $\tilde{\alpha}_j^{k}(\tau), \tilde{\phi}_j(\tau), \tilde{\Theta}(\tau)$ are defined in similar ways. Suppose the initial state is $|\psi(0)\rangle$, the wavefunction of the ideal final state and the final state under fluctuation are given by $U(\tau)|\psi(0)\rangle$ and $\tilde U(\tau)|\psi(0)\rangle$ respectively. The fidelity of the final state is given by 

\begin{align}
1-F&=1-|\langle\psi(0)|\tilde U(\tau)^\dag U(\tau)|\psi(0)\rangle|^2\\
&= 1-\left|\langle\psi(0)|\exp[i\sum_{j\in\{j_a,j_b\}}(\phi_j(\tau)-\tilde{\phi}_j(\tau))\sigma_j^x+i\left(\Theta(\tau)-\tilde{\Theta}(\tau)\right)\sigma_{j_a}^x\sigma_{j_b}^x   ]|\psi(0)\rangle\right|^2.
\end{align}
According to our pulse design scheme, $|\alpha_j^k(\tau)|$ is insensitive when pulse sequence satisfying Eq.~\eqref{eq:am}. So the contribution of infidelity mainly comes from the second term. The fidelity can be further estimated as 
\begin{align}
1-F&= 1-\left|\langle\psi(0)|\exp[i\left(\Theta(\tau)-\tilde{\Theta}(\tau)\right)\sigma_{j_a}^x\sigma_{j_b}^x  ]|\psi(0)\rangle\right|^2+O(\xi_{\omega}^4)\\
&\leqslant|(\Theta(\tau)-\tilde{\Theta}(\tau)|^2+O(\xi_{\omega}^4)\\
&=\Delta_{\omega}+O(\xi_{\omega}^4)
\end{align}
%The remaining task is to bound $|\Theta(\tau)-\tilde{\Theta}(\tau)|$.
where $\Delta_{\omega}=\left|\sum_k\frac{\partial\Theta(\tau)}{\partial \delta_k}\right|^2\xi_{\omega}^2$ is the error due to the fluctuation of motional mode frequencies. We have

\begin{align}
\left|\sum_k\frac{\partial\Theta(\tau)}{\partial \delta_k}\right|&= \left|\sum_{k} \eta_{k}^{2} b_{j_a}^{k} b_{j_b}^{k} \int_{0}^{\tau} {\rm d} t_{1} \int_{0}^{t_{1}} {\rm d} t_{2} 
\left(\Omega_{j_a}(t_1)\Omega_{j_b}(t_2)+\Omega_{j_b}(t_1)\Omega_{j_a}(t_2)\right) \cos \left[\delta_{k}\left(t_{1}-t_{2}\right)\right](t_1-t_2)\right|\\
&\leqslant\sum_k \eta_{k}^{2} b_{j_a}^{k} b_{j_b}^{k} \int_{0}^{\tau} {\rm d} t_{1} \int_{0}^{t_{1}} {\rm d} t_{2} \:
2\Omega_{\max}^2|t_1-t_2|\\
&\leqslant O(\eta_{\max}^{2}\Omega_{\max}^2\tau^3).
%&\leqslant\frac{1}{3} \eta_{\max}^{2}\Omega_{\max}^2\tau^3.
\end{align}
So $\Delta_\omega$ can be further bounded as 
\begin{align}\label{eq:Do}
\Delta_\omega\leqslant O(\eta_{\max}^4\Omega_{\max}^4\tau^6\xi_{\omega}^2).
%(1/9)\eta_{\max}^4\Omega_{\max}^4\tau^6\xi_{\omega}^2.
\end{align}
 Note that the Eq.~\eqref{eq:Do} is only an upper bound. In practice, the error may scales much slower.

With a similar method for mode frequency drift, we can also analysis the fluctuation errors for other parameters. Taking the Rabi frequency as an example, we assume that the Rabi frequency changes in the form of 
\begin{align}\label{eq:rabi_change}
\Omega_j(t)&\rightarrow \Omega_j(t)(1+\xi_{\Omega}).
\end{align}
We note that the noise model may be more complicated in practise, but the quasi-static approximation here is still valid in general. 
Firstly, in many color noise models, such as 1/f noise, the noise strength reduces with the noise frequency. As the quasi-static model corresponds to the lowest frequency, it can be a good approximation when low frequency components dominates. Moreover, even there are non-negligible high-frequency errors, the contribution to the infidelity is likely to be small due to the time-average effect.

Replacing the Rabi frequency in Eq.~\eqref{eq:evo} in the form of Eq.~\eqref{eq:rabi_change}, the infidelity satisfies $1-F=(\pi^2/4)\xi_{\Omega}^2+O(\xi_{\Omega}^4)$. Therefore, the protocol is also insensitive to the fluctuation of Rabi frequency.

\section{Simulation algorithm}\label{app:seq}
Below, by considering the commutation relation between spin and phonon degree of freedom, we developed an efficient simulation algorithm with linear runtime.
An important observation is that the Hamiltonian and Lindbladian terms in the master equation Eq.~\eqref{eq:mst} can be written as the sum of single mode terms,

\begin{align}
    H = \sum_k H_k,\quad \quad \mathbb{L}(\rho) = \sum_k \mathbb{L}_k(\rho)
\end{align}
where %\yx{[missing $j$ summation for $H_k$?]}
\begin{subequations}\label{eq:hlk}
\begin{align}
    H_k = &\sum_{j\in\{j_a,j_b\}}\Omega_{j}(t) \eta_{k} b_{j}^{k}\left(a_{k}^{\dagger} e^{i \delta_{k} t}+a_{k} e^{-i \delta_{k} t}\right) \sigma_{j}^{x},\\
    &\mathbb{L}_{k}(\rho)=\mathbb{L}_{k,\uparrow}(\rho)+\mathbb{L}_{k,\downarrow}(\rho)+\mathbb{L}_{k,\text{d}}(\rho)
    \end{align}
    and
    \begin{align}
\mathbb{L}_{k,\uparrow}(\rho) = &\Gamma_{k,\uparrow} \left(a_{k}^{\dagger} \rho(t) a_{k}-\frac{1}{2}\left\{a_{k} a_{k}^{\dagger}, \rho(t)\right\}\right)\nonumber,\\
    \mathbb{L}_{k,\downarrow}(\rho) =&\Gamma_{k,\downarrow} \left(a_{k} \rho(t) a_{k}^{\dagger}-\frac{1}{2}\left\{a_{k}^{\dagger} a_{k}, \rho(t)\right\}\right)\nonumber,\\
     \mathbb{L}_{k,\text{d}}(\rho) =&\Gamma_{k,\text{d}} \left(n_{k} \rho(t) n_{k}-\frac{1}{2}\left\{n_{k}^2, \rho(t)\right\}\right).
\end{align}
\end{subequations}
To better understand Eq.~\eqref{eq:hlk}, we may transfer it to the vectorized form,
\begin{equation}\label{eq:hlkv}
    \hat\rho(t) = \mathcal{T}   \exp \left[\int_0^\tau {\rm d}t\sum_k (\hat H_k (t) + \mathbb{\hat L}_k(t))\right]  \hat\rho(0)
\end{equation}
where $\mathcal{T}$ is time order operator, $\hat\rho$ is the density operator represented in the vector form~\cite{manzano2020short}. $\hat H_k $ and $\mathbb{\hat L}_k$ denote the
 Liouvillian operator in the matrix form with the correspondence $-i[H_k, \rho]\leftrightarrow\hat H_k\hat\rho$ and $\mathbb{ L}_k (\rho)\leftrightarrow \mathbb{\hat L}_k \hat\rho$.
%$\rho$ is the density operator, and $\hat H_k $ and $\mathbb{\hat L}_k$ respectively denote the
 %linear map on density matrix space with the correspondence $\hat H_k\rho = -i[H_k, \rho]$ and $\mathbb{\hat L}_k \hat\rho = \mathbb{ L}_k (\rho)$~\cite{manzano2020short,klimov2009group}. 
 It can be verified that when $k\neq k'$, we have  $[\hat H_k,\hat H_{k'}]=0$, $[\hat{\mathbb{L}}_k,\hat H_{k'}]=0$ and $[\hat{\mathbb{L}}_k,\hat{\mathbb{L}}_{k'}]=0$. Thanks to these commutation relations, evolution in Eq.~\eqref{eq:hlkv} can be simplified as the product of evolution under different modes  
\begin{equation}
    \hat\rho(t) = \prod_k \mathcal{T}  \exp \left[\int_0^\tau {\rm d}t \hat H_k (t) + \mathbb{\hat L}_k(t)\right] \hat\rho(0).
\end{equation}
In other words, the dissipation effect on each motional mode can be treated \textit{sequentially}. By transforming back to the matrix form, it can be verified that the solution of the final spin state (with motional modes traced out) of the master equation Eq.~\eqref{eq:mst} can be rigorously obtained by a sequential mode simulation described in Algorithm~\ref{alg:alg1}. It should be noticed that our algorithm relies on the decoupling of different phonon modes, which requires the system to be within the Lamb-Dicke regime.

\subsection{Simulation Time Complexity}
\label{ap:time}
As stated in the main text, direct simulation of a multiple-ion system with heating noise is intractable, as the dimension of the density matrix grows exponentially with respect to $N$. Nevertheless, the sequential simulation method introduced in section~\ref{sec:num_sim} overcomes this problem and serves as a powerful tool for simulating large ion crystal system. 

\begin{figure}[h]
    \centering
    \includegraphics[width = 9cm]{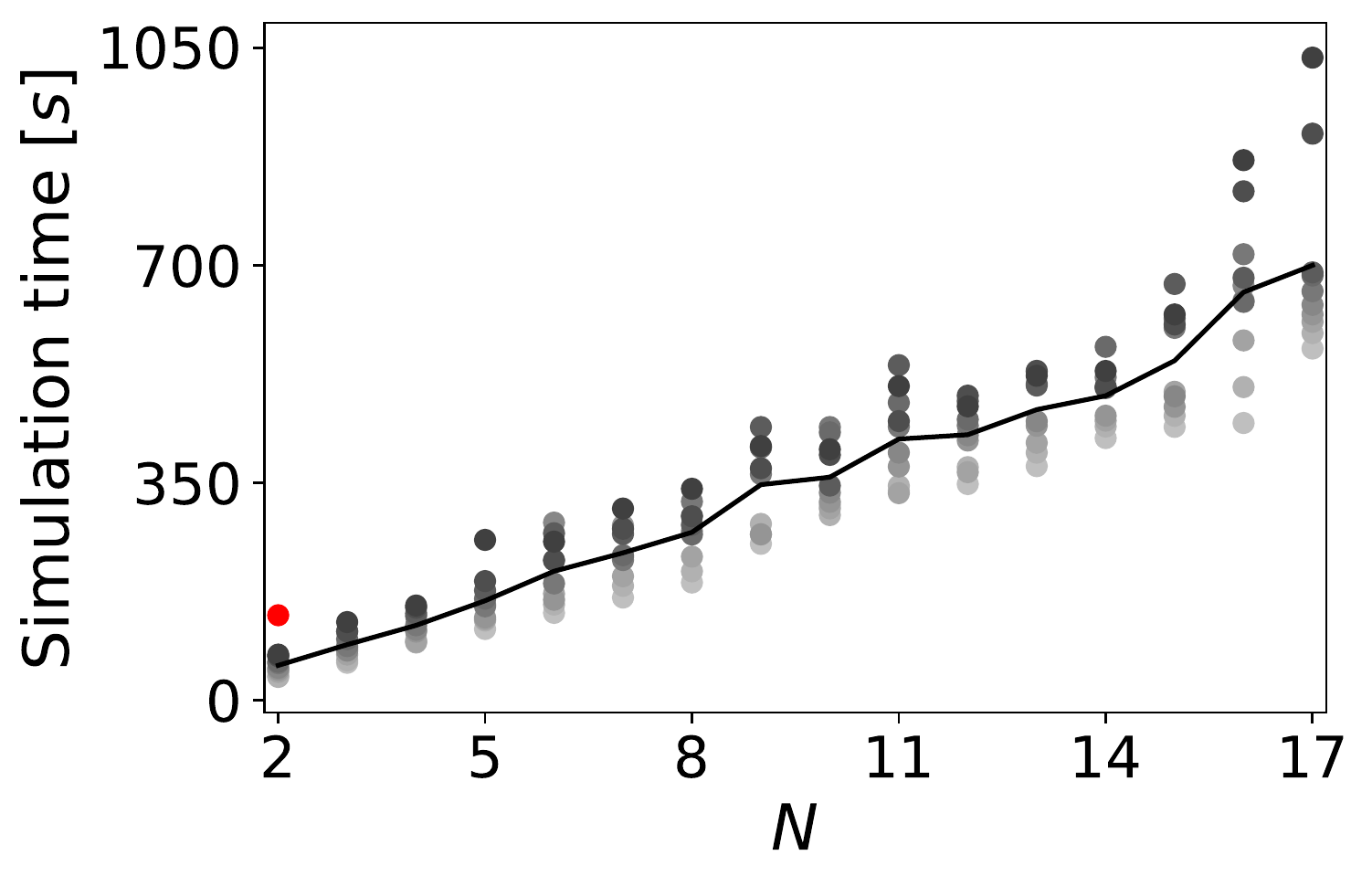}
       \caption{Total simulation time versus ion number $N$. Black dots: simulation time with Algorithm~\ref{alg:alg1}. Laser frequencies are set to be $\delta_{\min}=2\pi\times\{0.01{\rm MHz},0.02{\rm MHz},\cdots,0.10{\rm MHz}\}$, and the opacity of the dots increase with $\delta_{\text{min}}$. Black line: average over all laser frequencies.  
       Red dot: averaged simulation time of a brute-force simulation method for $N=2$. Simulation is executed on high-performance compting platform \textit{Weiming-No.1} at Peking University, Section C032M0256G, one node per task. Python version: python 3.8.13; Package version: qutip 4.6.3.}
       \label{fig:supp.simu_time}
\end{figure} 

Simulation time is illustrated in Fig.~\ref{fig:supp.simu_time}, with cut-off phonon number $N_c=10$ for all motional modes. We have applied \textit{mesolve} function from qutip 4.6.3, and the simulation is executed on High-performance computing platform \textit{Weiming-No.1} at Peking University. It can be noticed that the total time cost grows linearly with $N$, as expected from the theoretical analysis. In comparison, we also plot the averaged time cost of a brute-force simulation working on $N=2$, illustrated as the red dot in Fig.~\ref{fig:supp.simu_time}. We note that the $N=2$ is the only case we can work out using brute-force simulation. Even for $N=3$, the density matrix is too large to be simulated via \textit{mesolve} from qutip. 
\section{Supplementary Data and Figures}
\label{ap:sup}
%{\color{red}to be completed}
\begin{figure}[h]
    \centering
       \includegraphics[width = 16.4cm]{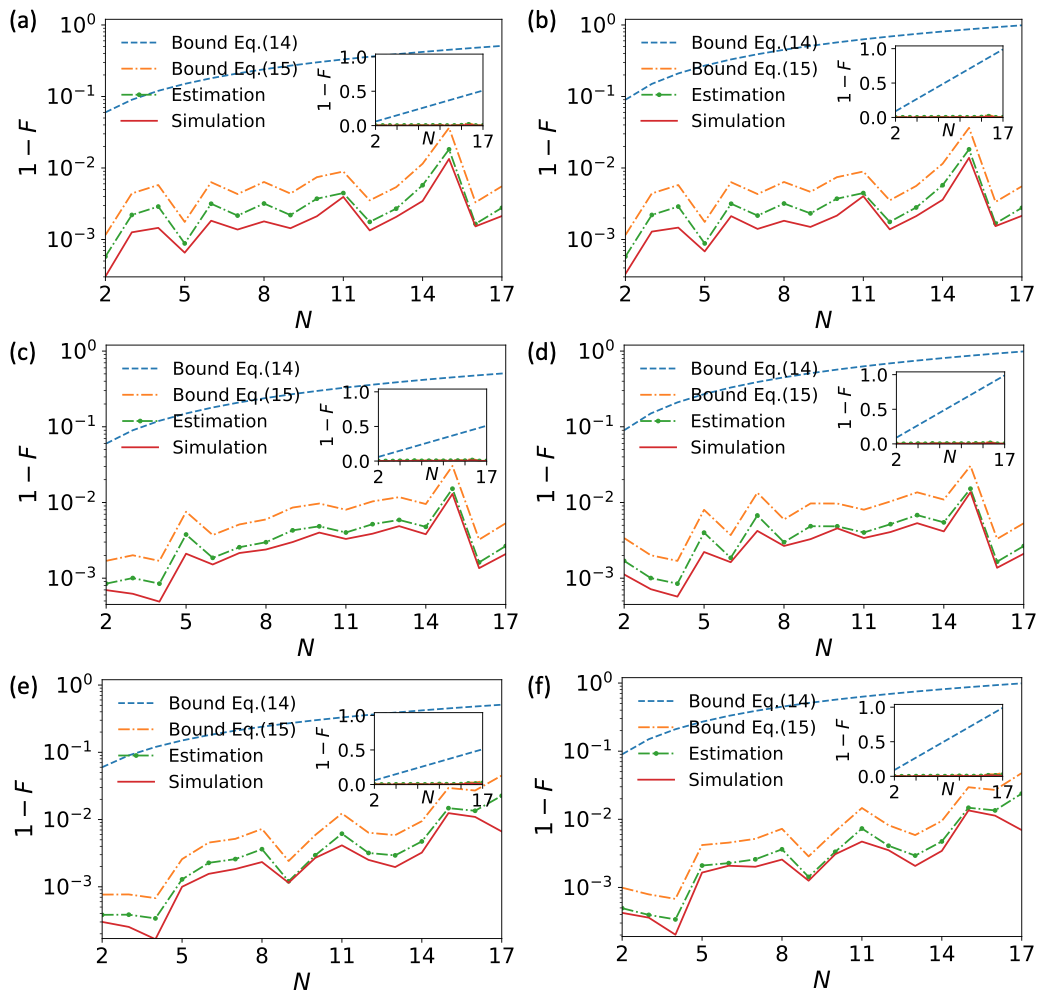}
       \caption{Upper bounds and simulation results for motional heating errors, illustrated for different laser pulses and noise models. The detuning gap $\delta_{\rm min}$ between the effective frequency of the laser and the smallest frequency of all collective modes are different: (a), (b) $\delta_{\min}=2\pi\times0.03{\rm MHz}$; (c), (d) $\delta_{\min}=2\pi\times0.06{\rm MHz}$; (e), (f) $\delta_{\min}=2\pi\times0.09{\rm MHz}$. The left subgraph on each row applies a uniform noise model, with the heating rate of every collective mode being identical, $\Gamma_{k,\uparrow}=\Gamma_{k,\downarrow} = 1$ phonons$/{\rm s}$; the right subgraphs apply the linear COM mode heating noise model with $\Gamma_{\rm COM} = N$ phonons$/$s and $\Gamma_{k\neq\rm COM,\uparrow}=\Gamma_{k\neq\rm COM,\downarrow} = 1$ phonons$/$s. We have set $\Gamma_{k,\text{d}}=0$ phonons$/$s for all subfigures.}
       %\label{fig:default}
       \label{fig:2_supp}% 
\end{figure} 

\end{appendix}
\end{document}